\documentclass[reprint,pre, bibnotes, nofootinbib, superscriptaddress,showpacs]{revtex4-1}
\pdfoutput=1
\usepackage{graphicx,subfigure}
\usepackage{graphicx,amsfonts}
\usepackage{amsmath}
\usepackage{verbatim}
\usepackage{amssymb}
\usepackage{bm}
\usepackage{amsthm}
\usepackage{textcomp}

\usepackage{natbib}
\usepackage{color}
\definecolor{darkblue}{rgb}{0,0,0.5}
\definecolor{darkred}{rgb}{0.5,0,0}
\usepackage[colorlinks=true,urlcolor=darkblue,citecolor=darkblue,linkcolor=darkred,hyperfootnotes=false]{hyperref}

\allowdisplaybreaks

\newcommand{\M}[1]{\mathcal{#1}}
\newcommand{\T}[1]{\text{#1}}

\newcommand{\erfc}{\text{erfc}}

\newcommand{\meas}[1]{{#1}^\text{m}}
\newcommand{\tmeas}[1]{{\tilde{#1}}^\text{m}}

\newcommand{\tl}[1]{\tilde{#1}}
\newcommand{\err}[1]{{#1}_{\text{e}}}
\newcommand{\tu}{U_{\text{trap}}}
\newcommand{\nn}{\nonumber}

\newcommand{\av}[1]{\left<#1\right>}
\newcommand{\xmol}{{x}_{\text{mol}}}
\newcommand{\cmol}{{X}_{\text{mol}}}
\newcommand{\ssn}{\sigma_{\text{SN}}^2}
\newcommand{\uncore}{E_{\text{u}}}
\newcommand{\uncorvar}{\sigma^2_{\text u}}

\newcommand{\be}{\begin{equation}}
\newcommand{\ee}{\end{equation}}
\newcommand{\bea}{\begin{eqnarray}}
\newcommand{\eea}{\end{eqnarray}}
\newcommand{\ba}{\begin{align}}
\newcommand{\ea}{\end{align}}

\begin{document}
 \title{Thermodynamic inference based on coarse-grained data or noisy measurements}
\author{Reinaldo Garc\'ia-Garc\'ia}
\email{reinaldomeister@gmail.com}
\affiliation{Laboratoire de Physique et M\'ecanique des Milieux H\'et\'erog\`{e}nes-UMR 
CNRS 7636, ESPCI, 10 rue de Vauquelin, 75231 Paris cedex 05, France}
\author{Sourabh Lahiri}
\affiliation{Laboratoire de Physico-Chimie Th\'eorique-UMR CNRS Gulliver 7083, PSL Research University, ESPCI, 
10 rue de Vauquelin, 75231 Paris cedex 05, France}
\author{David Lacoste}
\affiliation{Laboratoire de Physico-Chimie Th\'eorique-UMR CNRS Gulliver 7083, PSL Research University, ESPCI, 
10 rue de Vauquelin, 75231 Paris cedex 05, France}

\begin{abstract}
Fluctuation theorems have become an important tool in single molecule biophysics to measure free energy differences 
from non-equilibrium experiments. When significant coarse-graining or noise affect the measurements, the determination of the free energies becomes challenging. In order to address this thermodynamic inference problem, we propose improved 
estimators of free energy differences based on fluctuation theorems, which we test on a number of examples. 
The effect of the noise can be described by an effective temperature, which only depends on 
the signal to noise ratio, when the work is Gaussian distributed and uncorrelated with the error made on the work. 
The notion of effective temperature appears less useful for non-Gaussian work distributions or when the error is correlated with the work,
but nevertheless, as we show, improved estimators can still be constructed for such cases.
As an example of non-trivial correlations between the error and the work, we also consider measurements with delay, as described by 
 linear Langevin equations. 
\end{abstract}
\pacs{05.40.-a, 05.70.-a, 05.70.Ln} 

\maketitle
\section{Introduction}
\label{sec:intro}

Fluctuation theorems are symmetry relations, which constrain the probability distributions of thermody-
namic quantities arbitrarily far from 
equilibrium~\cite{Jarzynski-a,*Jarzynski-b,Crooks-a,*Crooks-b,Seifert-Review}. Their discovery has 
represented a major progress in our understanding of the second law of thermodynamics 
and has also accompanied many advances in the observation and manipulation of
various experimental non-equilibrium systems, such as biopolymers \cite{Ribezzi-Crivellari2014,Gupta}, manipulated colloids \cite{Wang,carberry}, 
mechanical oscillators or electronic circuits \cite{Ciliberto2010} or quantum devices \cite{Kueng2012}.

One major field of applications of fluctuations theorems lies in the determination of 
free-energies, through proper averaging of  
 the work within well-defined non-equilibrium ensembles.
In practice, in order to determine free energies using the Jarzynski relation~\cite{Jarzynski-a,*Jarzynski-b} 
for instance, a large number of experiments are required in order to ensure that
the rare trajectories which contribute the most are sampled correctly~\cite{jarzynski2006rare}. 

In addition to this sampling problem, other sources of errors in the determination of 
the free energy can arise from the measurement process itself. 
For instance, the experiment may involve some degrees of freedom which evolve on a much faster time scale 
than the response time of the measurement device, the experiment may not allow 
to measure all the degrees of freedom which are needed to evaluate the work or for some other reasons, 
the work is not properly evaluated from the measurements. Clearly, a difference can
easily arise between the true trajectories of the system and the coarse-grained or noisy trajectories, 
which are in fact recorded. This uncertainty in the trajectories leads to a difference between the true work and the measured work, 
which we call error and which limits our ability to determine free energy differences using fluctuation theorems.

In order to address this issue, a proper understanding of the way coarse-graining or 
measurement noise affects fluctuation relations is needed. 
The modifications of fluctuation relations due to coarse-graining have been studied by a number of authors 
following the original theoretical work of Rahav et al. \cite{Rahav2007} and motivated by various 
experimental systems such as
 manipulated colloids \cite{Tusch2014,Mehl2012}, granular systems \cite{Naert2012}, quantum dot devices \cite{BulnesCuetara2011,Kueng2012}, 
molecular motors \cite{pre-flashing-ratchet,Pietzonka2014}, and single biopolymer molecules \cite{Dieterich2015,Alemany_Inference,Ribezzi-Crivellari2014}. 
For instance, for molecular motors, the issue of coarse-graining is central, since only their 
position is typically available as a function of time experimentally. The chemical consumption of 
ATP from these molecules is hidden
and this limits our ability to use fluctuation theorems for molecular motors.
Naturally, for other systems, the precise modifications of the fluctuation relations will take various 
forms depending on the original dynamics and the 
way coarse-graining is performed \cite{Esposito2012_vol85,Bo2014,Michel2013}.

The present paper addresses the effect of coarse-graining or noise on fluctuation theorems of 
the Jarzynski and Crooks type.  
It is closely related to two recent studies, the first one on the error associated with finite 
time step integration in Langevin equations \cite{sivak2013using}
 and the second one on thermodynamic inference of free energy differences 
 in single molecules experiments \cite{Alemany_Inference,Ribezzi-Crivellari2014}.
Building mainly on these two works, we revisit this issue at a general level. We think that such an approach is  
pertinent since the question we are interested in is not bound to a specific experimental 
setup or dynamics: at some level, it originates from a fundamental 
property of entropy, namely its dependence on coarse-graining.

The remainder of the paper is organized as follows. In section \ref{sec:general}, we present general
properties of the correction factors to the Jarzynski and Crooks relations. Then in section \ref{sec:applications}, we 
first consider the simple case when the work and the error are Gaussian distributed and 
the error is uncorrelated with the work. This example is then extended in two ways: first by considering non-Gaussian work distributions
and then by considering the specific case that the error is linearly correlated with the work. 
We end in section \ref{sec:tweezers} by a numerical verification of our results based on specific choices of dynamics. 
This section also includes an analytical and numerical study of a model based on Langevin equations for which, correlations in the error arise 
 due to measurement delays.

\section{General properties of Fluctuation theorems with coarse-graining or noise}
\label{sec:general}

The Jarzynski relation~\cite{Jarzynski-a,*Jarzynski-b} allows to determine
equilibrium free-energy differences from an average of non-equilibrium measurements:
\begin{equation}
 \label{Jarzynski}
 \langle e^{-\beta W}\rangle_\Lambda=e^{-\beta\Delta F},
\end{equation}
where $W$ is the work done on a system and $\Lambda=\{\lambda(t)\}_{t=0}^\tau$ denotes a protocol of variation of a 
control parameter $\lambda(t)$ between time $0$ and time $\tau$, which starts initially
in an equilibrium state A corresponding to the value $\lambda_A=\lambda(0)$,
and ends up when the control parameter has reached $\lambda_B=\lambda(\tau)$ at time $\tau$.
Although the state reached by the
system at time $\tau$ is not in general an equilibrium one, $\Delta F=F_B-F_A$ represents the  
equilibrium free energy difference between states corresponding to $\lambda_A$ and $\lambda_B$. 
The average in Eq. (\ref{Jarzynski}), denoted by $\langle\ldots\rangle_\Lambda$, 
is taken over all non-equilibrium trajectories which are realized in this process. 

Very much related to the Jarzynski relation, the Crooks fluctuation theorem, 
constrains the ratio of probability distributions of the work associated with an arbitrary protocol which 
starts in an equilibrium state, $P(W)$, with respect to its time-reversed twin, $\tl{P}(W)$, associated with
$\tl{\Lambda}=\{\lambda(\tau-t)\}_{t=0}^\tau$~\cite{Crooks-a,*Crooks-b}:
\begin{equation}
 \label{Crooks}
 \ln\frac{P(W)}{\tl{P}(-W)}=\beta(W-\Delta F).
\end{equation}
Both, Eqs. (\ref{Jarzynski}) and (\ref{Crooks}) have been experimentally used to determine free-energy differences.
From Eq. (\ref{Jarzynski}) follows straightforwardly that $\beta\Delta F=-\ln\langle e^{-\beta W}\rangle$, while
from Eq. (\ref{Crooks}) one obtains $\beta\Delta F=\beta W_*$, where 
$W_*$ solves $P(W_*)=\tl{P}(-W_*)$.

As mentioned in the introduction, we are interested 
in situations in which the true work $W$ is not accessible due to coarse-graining or 
noise present in the measured variables or 
due to an incorrect evaluation of the work. 
To describe the first source of error, due to the trajectories, 
we distinguish the true trajectory of the system, $X=\{x(t)\}_{t=0}^\tau$
which will be typically inaccessible, from the measured (or coarse-grained) one which is 
accessible and which we shall denote by
$\meas{X}=\{\meas{x}(t)\}_{t=0}^\tau$. Unless we specify otherwise, 
the distribution of the initial condition of the 
true trajectory, namely $x(0)$, is assumed to be at equilibrium. In contrast, the distribution of the initial 
condition of the measured trajectory, namely $\meas{x}(0)$, 
does not need to be at equilibrium and is typically correlated with $x(0)$.

In order to describe the second source of error, at the level of the work itself, 
we assume that both works are evaluated from an Hamiltonian, but that two 
different Hamiltonians $H$ or $\meas{H}$ may be involved. More precisely, we 
define
\be
\label{true-work}
W[X] =\int_0^\tau dt\dot{\lambda}(t)\partial_\lambda H\big(x(t);\lambda(t)\big),
\ee
and
\be
\label{meas-work}
\meas{W}[X] =\int_0^\tau dt\dot{\lambda}(t)\partial_\lambda \meas{H}\big( \meas{x}(t);\lambda(t)\big).
\ee

With these notations, we write generally:
\be
\label{def-E}
W[X]=\meas{W}[\meas{X}] + E[\meas{X},X],
\ee
where $W[X]$ denotes the true value of the work defined for the true 
trajectory $X$, $\meas{W}[\meas{X}]$ is similarly the measured work 
associated with the measured (or coarse-grained) trajectory, and $E[\meas{X},X]$ is the corresponding error.
For simplicity, we choose not to indicate explicitly the dependence on
the driving $\Lambda$ in $W[X]$ and $\meas{W}[\meas{X}]$.
This error $E$ can frequently be modeled as a Gaussian distribution with non-zero mean and variance. 
Furthermore, it may in general depend on the duration of the experiment and on 
the rate of change of the driving protocol, although 
we can not exclude other contributions independent of the driving. 

Let us also introduce two corrections factors $R$ and $\Omega(\meas{W})$, which capture respectively the modifications of    
Eq. (\ref{Jarzynski}) and Eq. (\ref{Crooks}) due to measurement errors or coarse-graining.
The modified Jarzynski relation becomes
\begin{equation}
 \label{Jarzynski-exp}
 \langle e^{-\beta (\meas{W}-\Delta F)}\rangle_\Lambda=e^{\beta R},
\end{equation}
and the modified Crooks relation becomes
\begin{equation}
 \label{Crooks-exp}
 \ln\frac{\meas{P}(\meas{W})}{\tmeas{P}(-\meas{W})}=\beta \left( \meas{W}+\Omega(\meas{W})-\Delta F \right),
\end{equation}
where $\meas{P}(\meas{W})$ denotes the probability distribution of the measured
work values, which equals $\langle\delta(\meas{W}-\meas{W}[\meas{X}])\rangle_\Lambda$.
From these equations, it is apparent that both estimators of free energy are biased. Indeed, the first one leads 
to the estimate of free energy $\Delta\hat{F}=\Delta F-R\neq\Delta F$, 
while the second one leads to  
$\Delta\hat{F}_C=W_\star=\Delta F-\Omega({W}_\star)\neq\Delta F$,
where 
$W_\star$ solves $\meas{P}(\meas{W}_\star)=\tmeas{P}(-\meas{W}_\star)$.

To shorten the notations, we shall denote the symmetry functions as
\be
\label{Y}
Y(W) = \ln\frac{P(W)}{\tilde P(-W)},
\ee
and similarly
\begin{equation}
 \label{Ym}
 \meas Y(\meas W)=\ln\frac{\meas P(\meas W)}{\tmeas{P}(-\meas W)}.
\end{equation}

\subsection{A joint distribution function based formulation}

In order to evaluate the corrections factors $R$ and
$\Omega(\meas{W})$, we rely on a symmetry relation for joint distributions~\cite{us-unifying,us-joint}. 
To understand how it is derived, it is useful to recall 
that at the heart of Crooks relation, Eq. (\ref{Crooks}), there is a deeper statement on the
 path probability density of true trajectories which is
\begin{equation}
 \label{path-FT}
 \frac{\tl{\M{P}}[\tl{X}]}{\M{P}[X]}=e^{-\beta(W[X]-\Delta F)},
\end{equation}
where it has been assumed that the system's initial condition at $t=0$
corresponds to equilibrium.
The starting point of this derivation is the ratio of the joint probabilities of 
true and measured trajectories in the forward process to that in the reverse process:
 \begin{align}
   \frac{\M P[X,\meas X]}{\tl{\M{P}}[\tilde X,\meas{\tilde X}]} &= 
   \frac{\err{\M{P}}[\meas X|X]}{\err{\tl{\M{P}}}[\meas{\tilde X}|\tilde X]}\frac{\M{P}[X]}{\tl{\M{P}}[\tilde X]}\nn\\
   &= \exp[\beta(\meas W+E-\Delta F)+\err S],
   \label{joint-path-FT}
 \end{align}
with $\err S\equiv \ln(\err{\M{P}}[\meas X|X]/\tl{\err{\M{P}}}[\meas{\tilde X}|\tilde X])$ probing
the time reversal symmetry of the conditional probability $\err{\M{P}}[\meas X|X]$.
In the last step, we have used Eq.~\eqref{def-E} and Eq.~\eqref{path-FT}.
We can then write
\begin{align}
  \meas P(\meas W,E,\err S) &= \int \mathcal DX\mathcal D\meas X \M{P}[X,\meas X]\nn\\
  &\times \delta(\meas W-\meas W[\meas X])\delta(E-E[X,\meas X])\nn\\
  &\times \delta(\err S-\err S[X,\meas X]).
\end{align}
It is simple to show using Eq.~(\ref{true-work}) that the true work 
is antisymmetric under time reversal in the following sense: 
\be
W[X]=-\tl{W}[\tilde{X}],
\ee
where the tilde operation on $W$ or $\meas W$ indicates that dynamics occurs in the presence of a reversed protocol. 
Naturally, given the similarity of definitions between the true and the measured works,
the same property holds for the measured work:
\be
\meas{W}[\meas X]=-\tmeas{W}[\tmeas{X}],
\ee
As a result of these two relations, the error, defined in Eq.~\eqref{def-E}, is also antisymmetric under time reversal, 
$E[X,\meas X]=-\tl{E}[\tl X,\tmeas X]$. Then, using these relations and 
Eq.~\eqref{joint-path-FT} we get
\begin{align}
  \meas P(\meas W,E,\err S) &= e^{\beta(\meas W+E-\Delta F)+\err S}
  \int \mathcal DX\mathcal D\meas X \tl{\M{P}}[\tilde X,\meas{\tilde X}]\nn\\
  &\times \delta(\meas W+\tmeas W[\meas{\tilde X}])\delta(E+\tl{E}[\tilde X,\meas{\tilde X}])\nn\\
  &\times \delta(\err S+\err{\tl{S}}[\tilde X,\meas{\tilde X}])\nn\\
  &= e^{\beta(\meas W+E-\Delta F)+\err S}\meas{\tl P}(-\meas W,-E,-\err S).
\end{align}
Integrating over $\err S$, we have
\begin{align}
  \meas P(\meas W,E)\av{e^{-\err S}|\meas W,E}_{\Lambda} = e^{\beta(\meas W+E-\Delta F)}\meas{\tl P}(-\meas W,-E).
\end{align}
Therefore one finally arrives at the relation
\begin{equation}
 \label{error-ft-0}
 \ln\frac{\meas{P}(\meas{W},E)}{\tmeas{P}(-\meas{W},-E)}=\beta\big( \meas{W}+E-\Delta F+
 \Sigma(\meas{W},E)\big),
\end{equation}
where
\begin{align}
  \Sigma(\meas{W},E)\equiv -\ln\av{e^{-\err S}|\meas W,E}_{\Lambda}.
\end{align}

In the following, we restrict to the case where $\Sigma(\meas{W},E)=0$, which
holds when $\err{\M{P}}[\meas{X}|X]=\tl{\err{\M{P}}}[\tmeas{X}|\tl{X}]$.
As we shall see, this assumption is not too restrictive and allows already to derive some interesting results. 
Under this assumption, Eq. (\ref{error-ft-0}) simplifies to
\begin{equation}
 \label{error-ft}
 \ln\frac{\meas{P}(\meas{W},E)}{\tmeas{P}(-\meas{W},-E)}=\beta( \meas{W}+E-\Delta F),
\end{equation}
which is precisely the fluctuation theorem for the joint distribution of
the measured work and the error \cite{us-joint}.
From Eq. (\ref{error-ft})
 we can immediately derive Eq. (\ref{Jarzynski-exp})
\begin{equation}
 \label{corrected-Jarzynski-general}
 \langle e^{-\beta(\meas{W}-\Delta F)}\rangle_\Lambda=
\big\langle e^{-\beta E}\big\rangle_{\tl{\Lambda}}\equiv e^{\beta R},
\end{equation}
leading to the explicit form of the correction to the Jarzynski estimator:
\begin{equation}
 \label{R-general}
R =\beta^{-1} \ln \big\langle e^{-\beta E}\big\rangle_{\tl{\Lambda}}=\beta^{-1} \ln \int \tl{\rho}(E)e^{-\beta E}dE,
\end{equation}
in terms of the marginal time-reversed distribution of the error
\begin{equation}
 \label{marginal-error}
 \tl{\rho}(E)=\int \tmeas{P}(\meas{W},E)d\meas{W}
\end{equation}

We now proceed with Eq. (\ref{Crooks-exp}), which can be easily deduced from (\ref{error-ft}). We have:
\begin{align}
\label{Omega-derivation}
 \tmeas{P}(-\meas{W}) &=\int dE\tmeas{P}(-\meas{W},-E)\nonumber\\
 &=\int dE \meas P(\meas{W},E)e^{-\beta( \meas{W}+E-\Delta F)}\nonumber\\
 &=\meas P(\meas{W})e^{-\beta(\meas{W}-\Delta F)}
 \big\langle e^{-\beta E}|\meas{W}\big\rangle_{\Lambda}.
\end{align}
From Eq. (\ref{Omega-derivation}) we immediately obtain Eq. (\ref{Crooks-exp}) with the identification
\begin{equation}
 \label{Omega-def}
 \Omega(\meas{W})=-\beta^{-1}\ln \big\langle e^{-\beta E}|\meas{W}\big\rangle_{\Lambda}.
\end{equation}
A link between $\Omega(\meas W)$ and $R$ can be simply derived from the fact that the detailed
theorem Eq. (\ref{Crooks-exp}) must lead to the integral theorem Eq. (\ref{Jarzynski-exp}):
\begin{align}
 \label{closure}
 \langle e^{-\beta(\meas{W}-\Delta F)}\rangle_{\Lambda} &=e^{\beta \Delta F}
 \int d \meas{W} \meas P(\meas{W})e^{-\beta \meas{W}}\nonumber\\
 &=e^{\beta \Delta F}\int d\meas{W} \tmeas{P}(\meas{W})e^{\beta(\Omega(-\meas{W})-\Delta F)}\nonumber\\
 &= \langle e^{\beta\Omega(-\meas{W})}\rangle_{\tl{\Lambda}},
\end{align}
which implies after comparing with Eq. (\ref{Jarzynski-exp}):
\begin{equation}
 \label{closure-1}
 R=\beta^{-1}\ln \langle e^{\beta\Omega(-\meas{W})}\rangle_{\tl{\Lambda}}.
\end{equation}

Notice that $R$ only depends on the error distribution function in Eq.~(\ref{R-general}) or on the correlations between 
the measured work and the error in the equivalent formulation of Eq.~(\ref{closure-1}). In both cases, 
the true work does not explicitly
appears \cite{sivak2013using}. The same property holds for the correction $\Omega(\meas{W})$.

\subsection{Explicit corrections for uncorrelated error}
\label{sec:explicit}
In practice, the evaluation of the functions $R$ and $\Omega(\meas W)$ is rather 
difficult since this requires a knowledge of the joint 
distribution of the error and the measured work.
In order to progress, we introduce further assumptions in this section.

We can generally write the joint probability distribution of the measured
work and the error as
\begin{align}
  \label{simple JPDF}
  \meas{P}(\meas{W},E)d\meas W dE &=\meas{P}(\meas{W} | E) \rho(E)d\meas W dE\nonumber\\
  &= P(\meas{W}+E | E)\rho(E) dWdE\nn\\
  \Rightarrow \meas{P}(\meas{W},E)  &=P(\meas{W}+E | E) 
  \left| \frac{\partial W}{\partial \meas{W}}\right|_E \rho(E)\nonumber\\
                     &=P(\meas{W}+E | E) \rho(E),
\end{align}
where 
in the second line, we have changed variables from $\meas{W}$ to $W$ 
using Eq. (\ref{def-E}); this change of variable has a Jacobian unity since $E$ is fixed, hence the third line. 
When the error is uncorrelated with the true work, $P(\meas{W}+E | E)=P(\meas{W}+E )$, 
and we obtain the following factorization relation:
\be
\label{factorized form}
\meas{P}(\meas{W},E) = P(\meas{W}+E) \rho(E).
\ee
Thanks to the factorization property of Eq. (\ref{factorized form}), the
experimental work distribution becomes a simple convolution:
\begin{equation}
 \label{exp-convol}
 \meas{P}(\meas W)=\int dE P(\meas W+E){\rho}(E).
\end{equation}
Furthermore, the conditional probability of the work given the error is just the true
work distribution, but shifted, $\meas{P}(\meas W|E)=P(\meas W+E)$. By Bayes formula, the conditional probability
of the error given the work reads
\begin{equation}
 \label{cond-important}
 \meas{P}(E|\meas W)=\frac{P(\meas W+E){\rho}(E)}{\int dE P(\meas W+E){\rho}(E)}.
\end{equation}
From the last equation and (\ref{Omega-def}), we obtain the form of $\Omega(\meas W)$ 
in terms of the true work and the error distributions:
\begin{equation}
 \label{Omega-final}
 \Omega(\meas W)=-\frac{1}{\beta}\ln\frac{\int dE P(\meas W+E){\rho}(E)e^{-\beta E}}
 {\int dEP(\meas W+E){\rho}(E)}.
\end{equation}

Eqs. (\ref{exp-convol}) and (\ref{Omega-final}) constitute the first main result of the present paper.
These explicit expressions of the correction factors can be derived when it is possible to
 integrate out the contribution of the error independently
of the other degrees of freedom of the system. 
More precisely, we have used two main assumptions: the first one is the 
invariance under time reversal symmetry of $\err{\M{P}}[\meas{X}|X]$ and the 
second one is the statistical independence of $E$ and $W$.
As shown in Appendix~\ref{app:proof1}, taken together these assumptions
also imply the invariance of the error distribution under time-reversal symmetry, namely:
\be
\label{rho_E}
\rho(E)=\tl\rho(-E).
\ee
In the following, we present various applications of this
framework to specific work and error distributions.

\section{Consequences for specific work and error distributions}
\label{sec:applications}

\subsection{Uncorrelated Gaussian error and Gaussian work distribution}
\label{subsec:uncorr-Gaussian}

Before addressing more complex situations, it is instructive to consider
a simple case where the true work and error distributions are Gaussian, 
and the error is assumed to be uncorrelated with the true work, of
 mean $\varepsilon$ and of variance $\sigma^2$.
In this case, the experimental work distribution will also be a Gaussian, and the
correction factor to the Crooks fluctuation theorem, $\Omega(\meas W)$, will be a linear function of $\meas W$.
To be explicit, let us take the work and noise probability distributions of the form
\begin{align}
 \label{Gauss-work}
 P(W) &=\frac{1}{\sqrt{2\pi\sigma_W^2}}\exp\bigg[-\frac{(W-\langle W\rangle)^2}{2\sigma_W^2}\bigg],\\
 \label{Gauss-error}
 {\rho}(E) &=\frac{1}{\sqrt{2\pi\sigma^2}}\exp\bigg[-\frac{(E-\varepsilon)^2}{2\sigma^2}\bigg].
\end{align}
Naturally, since $W$ and $E$ are assumed to be uncorrelated, the variance  
of the measured work $\sigma_{\meas W}^2$ is simply the sum of the variances of the work  
and of the error: $\sigma_{\meas W}^2=\sigma_W^2+\sigma^2$.
Now, the bias in the Jarzynski estimator, $R$ can be evaluated using Eqs.~(\ref{R-general}), (\ref{rho_E}) 
and (\ref{Gauss-error}), with the result
\begin{equation}
 \label{Jarzynski-Gauss}
 R=\beta\frac{\sigma^2}{2}+\varepsilon,
\end{equation}
which depends on temperature, the variance of the noise and its mean.

Let us now calculate the bias in the Crooks estimator, $\Omega(\meas W)$ from Eqs. (\ref{Omega-final}), 
(\ref{Gauss-work}) and (\ref{Gauss-error}).
We find:
\begin{equation}
 \label{Omega-Gauss}
 \Omega(\meas W)=-\frac{\sigma^2}{\sigma^2+\sigma_W^2}
 \bigg(\meas W-\langle W\rangle+\beta\frac{\sigma_W^2}{2}-\varepsilon\ssn\bigg),
\end{equation}
where $\ssn=\sigma_W^2/\sigma^2$ is the signal-to-noise ratio.

\begin{figure}[t]
 \centering
 \includegraphics[scale=0.55]{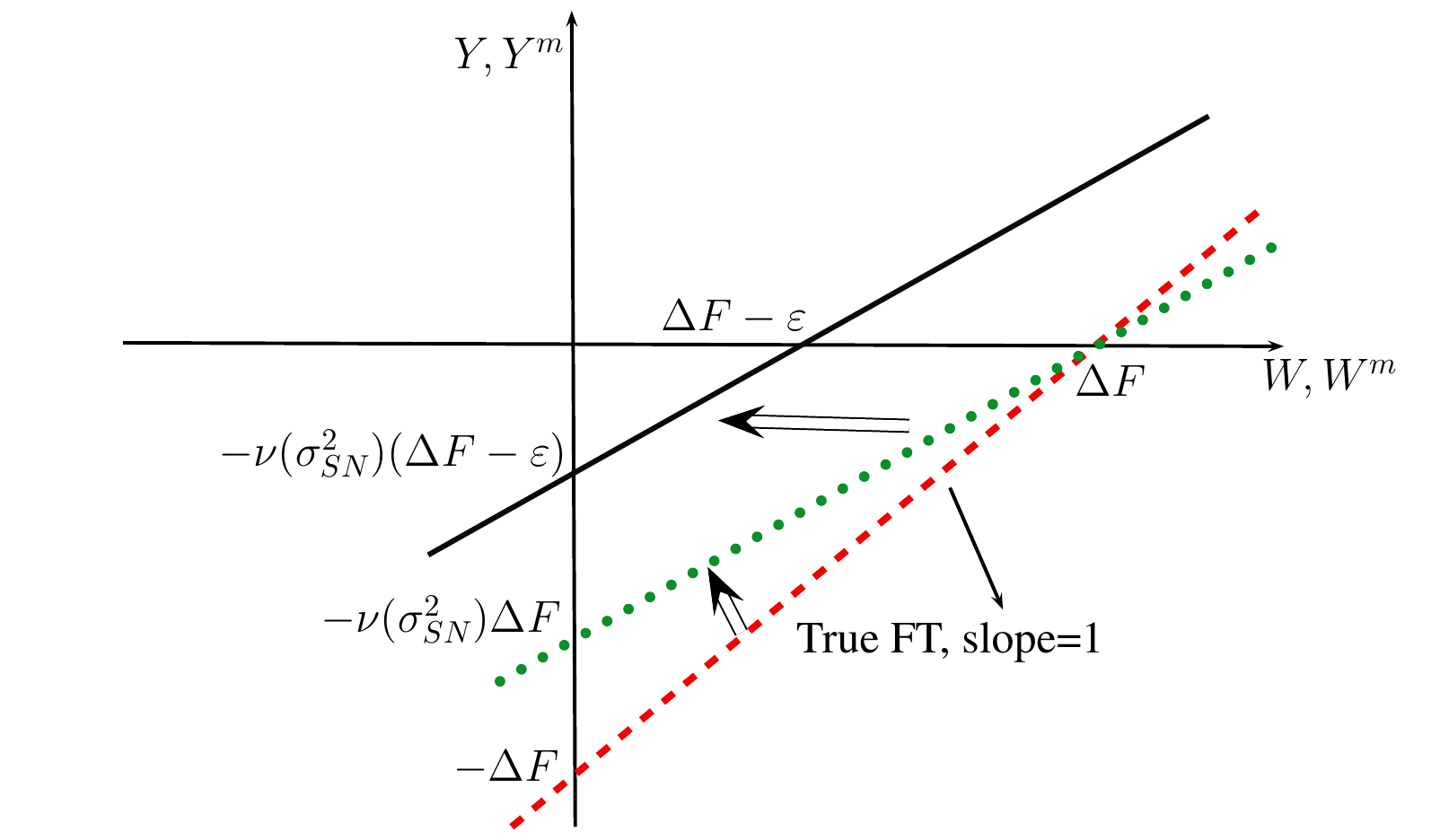}
   \caption{Sketch of the effect of Gaussian uncorrelated noise on symmetry functions, or equivalently on Crooks 
   fluctuation theorem, for a Gaussian
   work distribution. If $\varepsilon=0$, the measurement noise produces a decrease of the slope of 
   the symmetry function $\meas{Y}$ (green dotted line) as compared to $Y$ (red dashed line). This change of 
   slope (a rotation of the line) 
   does not affect the intersection point with the work axis, which corresponds to the free-energy difference, $\Delta F$. 
   When $\varepsilon\neq0$ however, the symmetry function should be in addition 
   translated by $\varepsilon$ (black solid line). 
   All energies are measured in units of $\beta^{-1}$.}   \label{fig1}
 \end{figure}
 
This result can be further simplified using the fluctuation theorem of the true work, namely 
$\langle e^{-\beta W}\rangle=e^{-\beta\Delta F}$, which is equivalent in this case to 
$\beta\sigma_W^2=2(\langle W\rangle-\Delta F)$. 
Thus, we obtain 
\be
\Omega(\meas W)=\vartheta(\ssn)(\meas W-\Delta F - \ssn \varepsilon),
\ee 
in terms of the signal-to-noise ratio
and the function 
 $\vartheta(y)=-1/(1+y)$.
In this simple case, the Crooks theorem for the distribution of the measured work reads
\begin{equation}
 \label{Exp-Crooks-Gauss}
\meas Y(\meas W)=\beta\nu(\ssn)(\meas W-\Delta F + \varepsilon),
\end{equation}
where $\meas Y$ is the symmetry function defined in Eq.~(\ref{Ym})
and $\nu(\ssn)$ is the function: 
\begin{equation}
 \label{fin}
 \nu(\ssn)=\frac{\ssn}{1+\ssn} = \frac{\sigma_W^2}{\sigma^2+\sigma_W^2}.
\end{equation}
As expected, the Crooks fluctuation theorem is recovered in the absence of noise,  
{\it i.e.} when $\varepsilon=\sigma=0$.

It is apparent with Eq.~(\ref{Exp-Crooks-Gauss}), that the mean of the error shifts the estimation of the 
free energy by a constant, while the variance of the error affects the slope of the symmetry function. 
When the mean of the error is zero ($\varepsilon=0)$, only the change of slope occurs. 
In that case, the Crooks estimator for the free-energy is not biased, while the Jarzynski estimator is. 
As the amount of noise or coarse-graining increases, the signal to noise ratio decreases, 
and the slope of the symmetry function decreases. Since the intersection point of this straight line 
with the work axis remains always equal to the free energy difference, the line  
undergoes a rotation with respect to the point $W=\Delta F$ on the work axis.
When the mean of the error is non-zero, this straight line undergoes
in addition an horizontal translation by the amount $\varepsilon$, as shown in Fig~\ref{fig1}.

Notice that the change of slope can be equivalently described by a change of temperature. 
One can thus introduce an effective temperature, equal to the temperature of the heat bath $T$ 
divided by $\nu(\ssn)$, therefore larger than $T$ since $\nu(\ssn) \le 1$ according to Eq. (\ref{fin}).
In the linear response regime, the same effective temperature will appear in the 
ratio of the response and correlation functions \cite{Verley2011_JSTATMech}.
It is important to appreciate however that this notion of effective temperature  
only applies to situations like the present one where the correction factor in the Crooks relation, 
namely, $\Omega(\meas W)$ is linear.
In general, this function is not linear as will become clear in the next examples 
and in the section reporting numerical results. In such cases, this effective 
temperature is less meaningful.

To summarize the results of this section, we have shown that an additive correction to the work due 
to an instrument error or noise
leads, in the case that the work and the error are Gaussian distributed, with uncorrelated error,
to a multiplicative factor for the temperature, in other words, to an effective temperature.
In addition, if the error has nonzero mean, the free-energy
estimator is shifted by an amount precisely equal to the mean value of the error.

\subsection{Uncorrelated Gaussian error with arbitrary work distribution}

We now show how to correct for measurement errors, when the true work distribution is arbitrary, 
keeping the same assumptions for the error (uncorrelated and Gaussian distributed).
We use Eq. (\ref{exp-convol}) in order to relate the probability distribution of the measured work 
to the probability distribution of the true work. 
Let us implement a shift by an arbitrary quantity $w$ in the argument of this distribution:
\begin{align}
 \label{correction-3}
\meas P(\meas W-w) &=\int dEP(\meas W+E-w)\rho(E),\\
 \label{correction-4}
 \tmeas{P}(-\meas W-w) &=\int dE\tl{P}(-\meas W-E-w)\tl{\rho}(-E)\nonumber\\
 &=e^{-\beta(\meas W-\Delta F+w)}\times\nonumber\\
 &\int dEP(\meas W+E+w)\rho(E)e^{-\beta E},
\end{align}
where we have used Eq.~(\ref{rho_E}) in the last step of Eq.~(\ref{correction-4}).

After the changes of variables $y=\meas W+E-w$ in (\ref{correction-3}) and $y=\meas W+E+w$ in
(\ref{correction-4}), we get:
\begin{align}
 \label{correction-5}
\meas P(\meas W&-w) =\int dy P(y)\rho(y-\meas W+w),\\
 \label{correction-6}
\tmeas{P}(-\meas W&-w) =e^{-\beta(\meas W-\Delta F)}\times\nonumber\\
 &\int dy P(y)\rho(y-\meas W-w)e^{-\beta(y-\meas W)}\nonumber\\
 &=e^{-\beta(\meas W-\Delta F)-2w\varepsilon/\sigma^2}\times\nonumber\\
 &\int dy P(y)\rho(y-\meas W+w)e^{-(\beta-2w/\sigma^2)(y-\meas W)},
\end{align}
where we have used, in the last step of Eq. (\ref{correction-6}), the explicit form of the error distribution,
Eq. (\ref{Gauss-error}).
It is now clear that choosing $w=\beta\sigma^2/2$
leads to:
\begin{equation}
 \label{amazing}
 \ln\frac{\meas P(\meas W-\beta\sigma^2/2)}{\tmeas{P}(-\meas W-\beta\sigma^2/2)}=\beta(\meas W-\Delta F+\varepsilon).
\end{equation}
Let us first analyze the case of unbiased error, $\varepsilon=0$.
We observe that, remarkably, the shift in Eq. (\ref{amazing}) removes the bias that 
was present in the Crooks estimator for measured work
and at the same time provides the correct slope for the fluctuation theorem. Thus, 
the transformation of Eq. (\ref{amazing}) solves in a simple way two problems at once: 
the need to calibrate the experiment against noise and 
the problem of the bias in the estimator. We shall illustrate this method using simulations in Sec. \ref{sec:uncorrelated_Gaussian}.

This result fully agrees with the results of Ref.~\cite{Ribezzi-Crivellari2014}, which is concerned with 
the inference of free-energies
 from partial work measurements in the context of single molecule experiments.  
The authors of this work showed that a shift of the type of Eq. (\ref{amazing}) can be used 
to exploit measurements of the ``wrong'' work in a symmetric dual trap system, in which one of the traps is fixed,
while the other one is moved. Such a transformation allows to recover the correct work distribution when the work 
distribution is Gaussian
and to eliminate the biases in the Jarzynski and Crooks estimators. 
However, as recognized by the authors, in the case of an asymmetric
setup of the traps, a shift of this kind does not permit to recover the correct work distribution 
(see Ref.~\cite{Ribezzi-Crivellari2014} for details).
This corresponds to our biased case, when $\varepsilon\neq0$. In such a case, the elimination
of the bias in the Crooks estimator is in principle not possible, at least not 
in the absence of additional information
on the error distribution \cite{Alemany_Inference}.

\subsection{Correlated Non-Gaussian error distribution}

Before moving to more complicated cases where the error is correlated with the true work 
and is non-Gaussian, let us consider a simple extension of the previous example. Let us assume that the error 
$E$ is of the form 
\begin{align}
\label{alpha-case}
  E=\alpha \meas W+\uncore,
\end{align}
so that the error is now the sum of a part which is proportional 
to the measured work, and another part $\uncore$, which is still 
uncorrelated with the true work $W$. By construction,  
the previous case is recovered for $\alpha=0$.
Note that when $W$ is non-Gaussian, this total error will also 
be non-Gaussian and correlated with $W$.

Let us introduce the probability distribution of the \emph{uncorrelated}
part of the error, $\rho_u(\uncore)$. As before with Eq. (\ref{simple JPDF}), we consider the joint distribution
\begin{align}
 \label{joint-alpha-0}
 \meas P(\meas{W},\uncore) &=\meas P(\meas{W} |\uncore)\rho_u(\uncore),\nn\\
&= P((1+\alpha)\meas{W}+\uncore |\uncore) \left| \frac{\partial W}{\partial \meas{W}} \right|_{\uncore}\rho_u(\uncore),\nn\\
&= (1+\alpha)P((1+\alpha)\meas{W}+\uncore | \uncore)\rho_u(\uncore).
\end{align}
Now, using the property that the variable $\uncore$ is uncorrelated with the true work $W$, we obtain
\begin{equation}
 \label{joint-alpha}
 \meas P(\meas W,\uncore)= (1+\alpha)P((1+\alpha)\meas W+\uncore)\rho_u(\uncore).
\end{equation}
An important point is that Eqs. (\ref{R-general}) and (\ref{Omega-def}) do not hold in terms of
$\rho_u(\uncore)$ and $\meas P(\meas W,\uncore)$ respectively, since $\uncore$ is not the total error,
but only its uncorrelated part. For instance, using Eq. (\ref{Omega-def}) and recalling that
$E=\alpha \meas W+\uncore$, we will now have:
\begin{align}
  \label{omega-alpha} 
  \Omega_\alpha(\meas W) &= -\frac{1}{\beta}\ln \av{e^{-\beta E}|\meas W}\nn\\
  &= \alpha\meas W-\frac{1}{\beta}\ln \av{e^{-\beta\uncore}|\meas W}\nn\\
  &\equiv\alpha\meas W+\Omega_{\alpha=0}\big((1+\alpha)\meas{W}\big),
\end{align}
where we have used the subscript $\alpha$ to make explicit the dependence on this parameter, and we have noticed,
given that $\uncore$ is uncorrelated from $W$, that the second term in the second line of Eq.~(\ref{omega-alpha})
is given exactly by Eq. (\ref{Omega-final}) with the substitution $\meas W\to(1+\alpha)\meas W$. 
Note that this result could also be derived by directly computing the joint probability of $\meas W$ and $E$, which
can be easily done as follows:
\begin{align}
\label{joint-transform}
\meas P(\meas W, E) &=\int d\uncore\meas P(\meas W,\uncore)\delta(E-\alpha\meas W-\uncore)\nn\\
 &=(1+\alpha)P(\meas W+E)\rho_u(E-\alpha\meas W),
\end{align}
where we have used Eq. \eqref{joint-alpha} to get the last line.
Thus, we have for $\meas P(\meas W)$
\begin{equation}
 \label{corr-meas-work}
 \meas P(\meas W)=(1+\alpha)\int dEP(\meas W+E) \rho_u(E-\alpha\meas W).
\end{equation}
Introducing $\meas P(E|\meas W)=\meas P(\meas W,E)/\meas P(\meas W)$, and using directly
Eq. (\ref{Omega-def}) together with Eqs. (\ref{joint-transform}) and (\ref{corr-meas-work}), we again
obtain (\ref{omega-alpha}).

Notice that, in particular,
when the distributions of the true work and $\uncore$ are Gaussian distributed, one obtains
\begin{align}
 \label{Omega-alpha-Gaussian}
 \Omega_\alpha(\meas W) &=\alpha\meas W+
 \vartheta(\ssn)\bigg[(1+\alpha)\meas W -\Delta F -\ssn\varepsilon\bigg]\nn\\
 &=\alpha\nu(\ssn)\meas W+\vartheta(\ssn)(\meas W-\Delta F-\ssn\varepsilon)\nn\\
 &\equiv\alpha\nu(\ssn)\meas W+\Omega_{\alpha=0}(\meas W).
\end{align}
with $\vartheta$, $\nu$, and $\ssn$ defined as before, but now in terms
of the variance of the \emph{uncorrelated} part of the error, $\sigma_u^2$.

It is worth noting, as we see from Eq.~(\ref{omega-alpha}), that this type of correlation only introduces a stretching
of the original $\Omega_{\alpha=0}$ via a rescaling of $\meas W$, 
plus an additional correction which is \emph{linear} in $\meas W$. In particular, in the Gaussian
case the stretching can be reabsorbed in the linear correction because $\Omega$ is linear in $\meas W$ for $\alpha=0$.

For the case of non-Gaussian work distributions but with a Gaussian distribution 
of $\uncore$, it is interesting to seek a relation 
of the type of Eq.~\eqref{amazing} as improved estimators of free energy.
Proceeding in the same way as before, the expressions for the forward 
and reverse probability distributions of the measured work shifted by an amount $w$ are:
\begin{align}
  \label{s-f}
  \meas{P} &(\meas W-w) = \int d\uncore \meas{P} (\meas W-w,\uncore)\nn\\
  &= (1+\alpha)\int d\uncore ~P\big((1+\alpha)(\meas W-w)+
  \uncore\big)\rho_u(\uncore), 
\end{align}
and
\begin{align}
  \label{s-r}
  \tmeas{P} &(-\meas W-w) = \int d\uncore \tmeas{P} (-\meas W,-\uncore)\nn\\
  &= (1+\alpha)e^{-\beta((1+\alpha)(\meas W+w)-\Delta F)}\nn\\
  &\times\int d\uncore ~P\big((1+\alpha)(\meas W+w)+\uncore\big)e^{-\beta\uncore}\rho_u(\uncore),
\end{align}
where we have used the relation
\be
\label{uncore-symmetry}
\rho_u(\uncore)=\tl \rho_u(-\uncore),
\ee
which holds under the same assumptions leading to Eq.~\eqref{error-ft}, 
as shown in Appendix~\ref{app:proof2}.

Let us now assume $\rho_u(\uncore)$ has a mean $\varepsilon$ 
and a variance $\uncorvar$, and for any arbitrary $w$, let us introduce the 
shifted symmetry function
\begin{align}
  \meas Y_{\text{sh}}(\meas W,w)=\ln\frac{\meas P(\meas W-w)}{\tmeas{P}(-\meas W-w)}.
  \label{Y_sh}
\end{align}
It can be shown that when $w=w^*=\frac{\beta\uncorvar}{2(1+\alpha)}$, 
this shifted symmetry function has a simple form:
\begin{align}
  \meas Y_{\text{sh}}(\meas W,w^*)=  \beta\bigg((1+\alpha)\meas W-\Delta F+\varepsilon\bigg).
  \label{amazing1}
\end{align}

It is important at this point to contrast this result with that obtained in Eq.~\eqref{amazing} 
for $\alpha=0$.
Although one obtains again a linear relation for the shifted symmetry function, the slope is not 
one (in units of $k_B T$) but $1+\alpha$. Since a priori neither $\varepsilon$ nor $\uncorvar$ are
known, one should vary the shift parameter $w$ in a plot of $ \meas Y_{\text{sh}}(\meas W,w)$ versus $\meas{W}$, 
until the data points collapse on a straight line. From the value of the slope 
of that line, the value of $\alpha$ can be inferred, and from 
the actual value of $w^*$, the value of $\uncorvar$ can then be deduced.
To apply this method, it is important to be sure that there is 
a unique value of the optimal shift $w^*$. We adress this point in appendix \ref{app:proof} by 
proving that indeed there is a unique optimal shift and furthermore that for no other value of 
$w$, the symmetry function is a linear function of $\meas{W}$. Naturally, this proof includes the 
case $\alpha=0$ considered previously.

When $\varepsilon=0$, this transformation of the symmetry function leads to a complete calibration since no other parameter
needs to be fixed, and the correct estimate of the free-energy difference can be recovered, as we shall 
illustrate numerically in Sec. \ref{sub-corr-err}.
However, when $\varepsilon\neq0$, the estimator is biased by the mean of the error in a way which can not be fixed 
in the absence of additional information, as also found in the previous case.

\section{Applications to specific choices of dynamics for the measured variable}
\label{sec:tweezers}

In this section, we shall apply the theoretical framework developed in previous sections to some specific 
dynamics for the measured variable. 
Before we do so, we discuss the choice of measured variables 
in single molecule experiments (typically position or force).
Then, assuming the position is the measured variable, we discuss the consequences of the particular choice 
of the relation between the dynamics of the measured position and that of the true position. 
Here, we shall restrict ourselves to two separate cases:

(a) Simple additive noise: the measured position $\meas x$ and the true position $x$ are related by
\be
\label{caseA}
  \meas x(t)=x(t)+\eta(t),
\ee

(b) Additive noise with delay: the measured position $\meas x$ and the true position $x$ are related by
\be
\label{caseB}
  \tau_r\meas{\dot x} = x-\meas x + \eta(t).
\ee
From an experimental point of view, case $(a)$ describes purely random measurement 
errors, which corresponds to the assumption that $\eta$ and $x$ are uncorrelated.
In contrast, case $(b)$ describes a case where these variables are correlated 
because the measurement device introduces a delay between $x(t)$ 
and its measured value, $\meas x(t)$.
Clearly, both cases are relevant experimentally.

Furthermore, for both dynamics $(a)$ and $(b)$, we assume the distribution of $x(0)$ to be an equilibrium one, 
while that of $\meas x(0)$ is not, but corresponds to a stationary non-equilibrium distribution. 
The system can be prepared in such a state at $t=0$ by starting the evolution at a time $t=-\infty$ 
in the absence of driving, so that the distributions of $x(0)$ and $\meas x(0)$ are both stationary. Naturally, both  
variables $x(0)$ and $\meas x(0)$ may still be correlated with each other.

\subsection{Choice of measured variable: position vs. force}

Before implementing the above dynamics, let us now discuss a practical question 
regarding the choice of measured variables 
in single-molecule experiments.  
In a first setup, where the position is measured, the Hamiltonian which is typically used has the form: 
$H(\xmol,x;\lambda)=H_{\T{mol}}(\xmol)+H_{\T{coup}}(\xmol,x)+\tu(x;\lambda)$, where
$H_{\T{mol}}(\xmol)$ describes the macromolecule under study 
(a DNA filament or RNA hairpin, for instance), with $\xmol$ labeling the relevant degrees of freedom of that system. 
This molecule is attached to a bead which is held in an optical trap, and the energy of the bead 
is given by $\tu(x;\lambda)$, where $x$ is the position of the bead and $\lambda$ the position of the trap center. 
Finally, $H_{\T{coup}}(\xmol,x)$ accounts for the coupling between
the molecule and the bead.

Usually, the calibration of optical tweezers relies on a harmonic approximation for the trapping potential,
$\tu(x;\lambda)=\kappa(x-\lambda)^2/2$, where $\kappa$ denotes the stiffness of the trap and $\lambda$ the position
of its center.
In this case, the work is
\begin{align}
 \label{Work}
 W[X] &=\int_0^\tau dt\dot{\lambda}(t)\partial_\lambda H\big(\xmol(t),x(t);\lambda(t)\big)\nonumber\\
 &=\int_0^\tau dt\dot{\lambda}(t)\partial_\lambda U_\text{trap}\big(x(t);\lambda(t)\big)\nonumber\\
 &=\kappa\int_0^\tau dt\dot{\lambda}(t)\big[\lambda(t)-x(t)\big],\nonumber
\end{align}
which does not depend explicitly on the degrees of freedom of the molecule under study characterized by $\cmol=\{\xmol(t)\}_{t=0}^\tau$. 
In this case, the work on the system is exactly equal to the work on the bead, since the trap is the only term of the Hamiltonian which depends on
$\lambda$.
The structure of the error in this situation is very simple:
\begin{equation}
 \label{work-decomposition-harmonic}
 E[X, \meas{X}]= \kappa\int_0^\tau dt\dot{\lambda}(t)\big[\meas{x}(t)-x(t)\big],
\end{equation}
which shows that the error increases with the driving speed $\dot\lambda$ and accumulates 
with the duration of the experiment $\tau$.

One limitation of such a setup where the position is measured lies 
in the harmonic approximation used for the trapping potential, 
an approximation which is expected to fail at large distances 
from the bead to the center of the trap. Furthermore, recent studies
have found great variability in the trap stiffness as a function of the position, even in the region where a constant
stiffness was expected~\cite{Jahnel:11}. To overcome such issues, 
a different setup is often preferred, where no assumption 
on the form of the trapping potential is needed.

In this alternative setup, the force rather than the position, is directed measured from the change in the momentum flux 
of the light beam impinging on the optical trap \cite{Smith2003}.
There is no need to assume a particular form of the trapping potential: 
one rather measures the force signal, $\meas{f}(t)$, which also has some noise
(i.e., $\meas{f}\neq f$, the true force exerted by the optical trap). 
The position of the center of the trap is the control parameter which we assume to be error free as we did so far. 
In some setups one does not have direct access to the position of the trap and one has
also to infer it with some error, but we dismiss that possibility here and assume that
this is our control parameter~\footnote{In certain 
experiments one fixes the force letting
the trapping velocity free. In that case a feedback mechanism is necessary. 
We do not address that case here.}.
For this setup the work reads:
\begin{equation}
 \label{work-exp}
 W[\M F]=\int_0^\tau\dot{\lambda}(t)f(t)dt,
\end{equation}
with $\M F=\{f(t)\}_{t=0}^\tau$. Note that the trapping potential $\tu(x;\lambda)\equiv\tu(x-\lambda)$, 
thus, $f=-\partial_x\tu
=\partial_{\lambda}\tu$, and the definition (\ref{work-exp}) coincides 
with the Jarzynski work~\cite{Jarzynski-a,*Jarzynski-b}, 
satisfying the nonequilibrium work theorem in the form given by (\ref{Jarzynski}). 
In this case the structure of the error is also very simple
\begin{equation}
\label{work-decomposition-force}
 W[\M F]=\meas W[\meas{\M F}]-\int_0^\tau dt\dot{\lambda}(t)\big[\meas{f}(t)-f(t)\big]. 
\end{equation}
It is worth noting that both, Eq. (\ref{work-decomposition-harmonic}) and
Eq. (\ref{work-decomposition-force}), have the same structure. In addition, note that
the assumption that $\lambda$ is error free is not very dangerous. This can be seen as follows.
In the first setup, one can redefine the distances and consider the error in measuring $\lambda-x$
instead of $x$ alone. In the second case, one does not need to know the value of $\lambda$ in
order to calculate the work because the force is directly recorded. 
In both cases what remains free is the pulling velocity, $\dot{\lambda}$,
which is very well controlled even if $\lambda$ itself is not.

Since it is a rather simple matter to switch between notations for the force setup and 
that for the position setup, we limit ourselves in the rest of the paper to only one case, 
which we chose to be the position setup.

\subsection{Corrected Jarzynski estimator}
\label{subsec:Jarzynksi}
Let us derive the correction to the Jarzynski estimator in the presence of measurement error $\eta$
within dynamics $(a)$ defined in Eq.~(\ref{caseA}).
\begin{align}
 \label{bias-work}
\langle e^{-\beta \meas{W}}\rangle_\Lambda
= &\int\M{D}\meas{X}\meas{\M{P}}[\meas{X}] e^{-\beta\meas W[\meas{X}]}\nonumber\\
= &\int\M{D}\meas{X}\M{D}X\err{\M{P}}[\meas{X}|X]\M{P}[X] e^{-\beta\meas W[\meas{X}]}\nonumber\\
= &\int\M{D}X\M{P}[X] e^{-\beta W[X]}\big\langle e^{\beta E[\meas{X},X]}
\big|X \big\rangle_\Lambda.
\end{align}
In this case one has 
$\err{\M P}[\meas{X}|X]=\M{P}[\pmb{\eta}|X]\equiv \M{P}_\eta[\pmb{\eta}]$, where $\M{P}_\eta$
is the path probability density of the error trajectory $\pmb{\eta}=\{ \eta(t) \}_0^{\tau}$. 
Thus, since the error in Eq. (\ref{work-decomposition-harmonic}) is a linear functional
of $\pmb{\eta}$, it can be integrated explicitly. We thus have
\begin{align}
 \label{bias-work-2}
 \langle e^{-\beta\meas W}\rangle_\Lambda
 &=\big\langle e^{-\beta W[X]}\big\rangle_\Lambda e^{K_\eta[\kappa\beta\dot{\lambda}]}\nonumber\\
 &\equiv
e^{-\beta(\Delta F-(1/\beta)K_\eta[\kappa\beta\dot{\lambda}])},
\end{align}
where we have used the Jarzynski equality, Eq. (\ref{Jarzynski}),
and we have introduced the generating functional of the cumulants of $\M{P}_\eta$,
$K_\eta[J]=\ln\int\M{D}\pmb{\eta} \M{P}_\eta[\pmb{\eta}]\exp[\int dtJ(t) \eta(t)]$. From this, we obtain 
the following estimate of the free energy, $\Delta\hat{F}(N)$ as:
\begin{equation}
 \label{main-result-1}
\Delta\hat{F}(N)=\Delta F-\frac{1}{\beta}K_\eta[\kappa\beta\dot{\lambda}].
\end{equation}
As stated before, the bias in the estimation of the free-energy difference only depends 
on the statistical properties of the error associated to measurement apparatus.
This result is fully compatible with the expression of the correction $R=k_B T K_\eta[\kappa\beta\dot{\lambda}]$ 
obtained from Eq. (\ref{R-general}) 
when $\rho(E)$ is assumed to be symmetric under time-reversal symmetry. 

Let us discuss now the validity of the factorization property 
Eq. (\ref{factorized form}), or equivalently, of the convolution formula of Eq. (\ref{exp-convol}). 
To illustrate this, let us consider a simple case where the optical trapping is assumed
to be parabolic, whereas in reality, it is not. In that case, the measured work is
\begin{align}
 \label{factor-fail-1}
 \meas W &=\meas \kappa\int_0^\tau dt\dot{\lambda}(t)[\lambda(t)-\meas x(t)]\nonumber\\
 &=\meas \kappa\int_0^\tau dt\dot{\lambda}(t)[\lambda(t)-x(t)-\eta(t)]
\end{align}
while the true work reads
\begin{equation}
 \label{factor-fail-2}
 W=\int_0^\tau dt\dot{\lambda}(t)\partial_\lambda U_{\text{trap}}(\lambda-x),
\end{equation}
where we have introduced the \emph{experimental} stiffness of the trap, $\meas \kappa$, and
the dynamics of Eq.~(\ref{caseA}) with $\eta(t)$ uncorrelated with $x(t)$.
Indeed, the error at time $\tau$, $E=W-\meas W$, is in general correlated with the values of $W$ not only
at time $\tau$, but even at earlier times. This is very easy to see by noting, from Eq.~(\ref{factor-fail-1}), that
we have $\lambda(t)-x(t)=\meas{\dot{W}}(t)/\meas \kappa\dot{\lambda}(t)+\eta(t)\equiv
(\meas \kappa\dot{\lambda}(t))^{-1}[\dot{W}(t)-\dot{E}(t)]+\eta(t)$. Substituting this back in (\ref{factor-fail-2})
we clearly see that $W$ and $E$ are in general correlated in a highly non-local way in time even in this simple case,
so that Eqs. (\ref{factorized form}) and (\ref{exp-convol}) do not hold anymore.

It is worth noting, however, that there is a particular case where one can still make the assumption 
that correlations are local in time.
This happens when the true trapping potential is still parabolic, but the stiffness is not correctly estimated, its value is 
$\kappa\neq\meas \kappa$. It is
easy to see that in this case we have
\begin{align}
 \label{factor-fail-3}
 W &=\kappa\int_0^\tau dt\dot{\lambda}(t)[\lambda(t)-x(t)]\nonumber\\
 &=\kappa\int_0^\tau dt\dot{\lambda}(t)[\lambda(t)-\meas x(t)+\eta(t)]\nonumber\\
 &=\frac{\kappa}{\meas \kappa}\meas \kappa\int_0^\tau dt\dot{\lambda}(t)[\lambda(t)-\meas x(t)]+
 \kappa\int_0^\tau dt\dot{\lambda}(t)\eta(t)\nonumber\\
 &\equiv\frac{\kappa}{\meas \kappa}\meas W +\uncore,
\end{align}
where $\uncore=\kappa\int_0^\tau dt\dot{\lambda}(t)\eta(t)$ 
is still uncorrelated from $W$, while the error $E=W-\meas W=[(\kappa/\meas \kappa)-1]\meas W+
\uncore$ is not. 
This shows that miscalibration of the trap stiffness introduces a correlated error of the form 
considered before in Eq.~(\ref{alpha-case}), 
with $\alpha$ given by $\alpha=\kappa/\meas \kappa-1$. 

\subsection{Fraction of second-law violating trajectories in terms of measured work}

Clearly, the fraction of trajectories that transiently violate the second 
law is different for the true and the measured works. 
For Gaussian distributed work, this fraction is analytically calculable, 
following the method of \cite{Sahoo2011}. We begin with the relation
\begin{align}
  & \int d\meas W P(\meas W)e^{-\beta\meas W} = e^{-\beta(\Delta F-R)}\nn\\
  &=\frac{1}{\sqrt{2\pi\sigma_{\meas W}^2}}\int d\meas W 
  \exp\bigg[-\frac{(\meas W-\av{\meas W})^2}{2\sigma_{\meas W}^2}-\beta \meas W\bigg].
\end{align}
A simplification of this relation leads to
\begin{align}
  \frac{1}{2}\beta\sigma_{\meas W}^2 = \av{\meas W}-\Delta F+R,
  \label{FDT}
\end{align}
where $R$ is given by Eq. \eqref{Jarzynski-Gauss}.
Now we can readily calculate the fraction of atypical trajectories (i.e. the ones that 
transiently violate the second law) by integrating the work probability distribution from $-\infty$ to $\Delta F$:
\begin{align}
\meas{\mathfrak{f}} &= \frac{1}{\sqrt{2\pi\sigma_{\meas W}^2}}\int_{-\infty}^{\Delta F}d\meas{W}
  \exp\bigg[-\frac{(\meas W-\av{\meas W})^2}{2\sigma_{\meas W}^2}\bigg]\nn\\
  &= \frac{1}{2}\erfc\bigg[\frac{\av{\meas W}-\Delta F}{\sqrt{2\sigma_{\meas W}^2}}\bigg].
\end{align}
Let us assume that the error in the measured work has non-zero mean $\varepsilon$. Using the definition of measured work and the 
fluctuation theorem for true work, we have: 
\begin{align}
  \av{W} &= \av{\meas W}+\varepsilon = \Delta F+\frac{1}{2}\beta\sigma_W^2.
\end{align}
Using \eqref{FDT}, one then obtains
\begin{align}
\meas{\mathfrak{f}}  &= \frac{1}{2}\erfc\bigg[\frac{1}{2}\frac{\av{\meas W}-\Delta F}{\sqrt{\beta^{-1}
  (\av{\meas{W}}-\Delta F+R)}}\bigg]\nn\\
&= \frac{1}{2}\erfc\left[\frac{1}{2}\frac{\frac{1}{2}\beta\sigma_W^2-\varepsilon}{\sqrt{\frac{1}{2}(\sigma_W^2+\sigma^2)}}\right].
\label{fraction}
\end{align}
We note that 
\begin{align}
  \meas{\mathfrak{f}}  &= \frac{1}{2}\erfc\bigg[\frac{\beta}{2\sqrt 2}~\frac{\sigma_W^2-2\beta^{-1}\varepsilon}{\sqrt{\sigma_W^2+\sigma^2}}\bigg]\ge \frac{1}{2}\erfc\bigg[\frac{\beta}{2}\sqrt{\frac{\sigma_W^2}{2}}\bigg]=\mathfrak{f},
\end{align}
if $\varepsilon\ge 0$. In that case, measurement errors cause an overestimation of the fraction of trajectories transiently violating the second law.
Furthermore, note that $\meas{\mathfrak{f}}>1/2$ only if the argument of the error function is negative. 
Thus, we will observe apparent violations of the second law if the error is positive and sufficient large so that 
$\varepsilon>\beta\sigma_W^2/2$.
In this case, the mean of the measured work is less than $\Delta F$. 

\begin{figure}[!h]
  \centering
  \includegraphics[width=8cm]{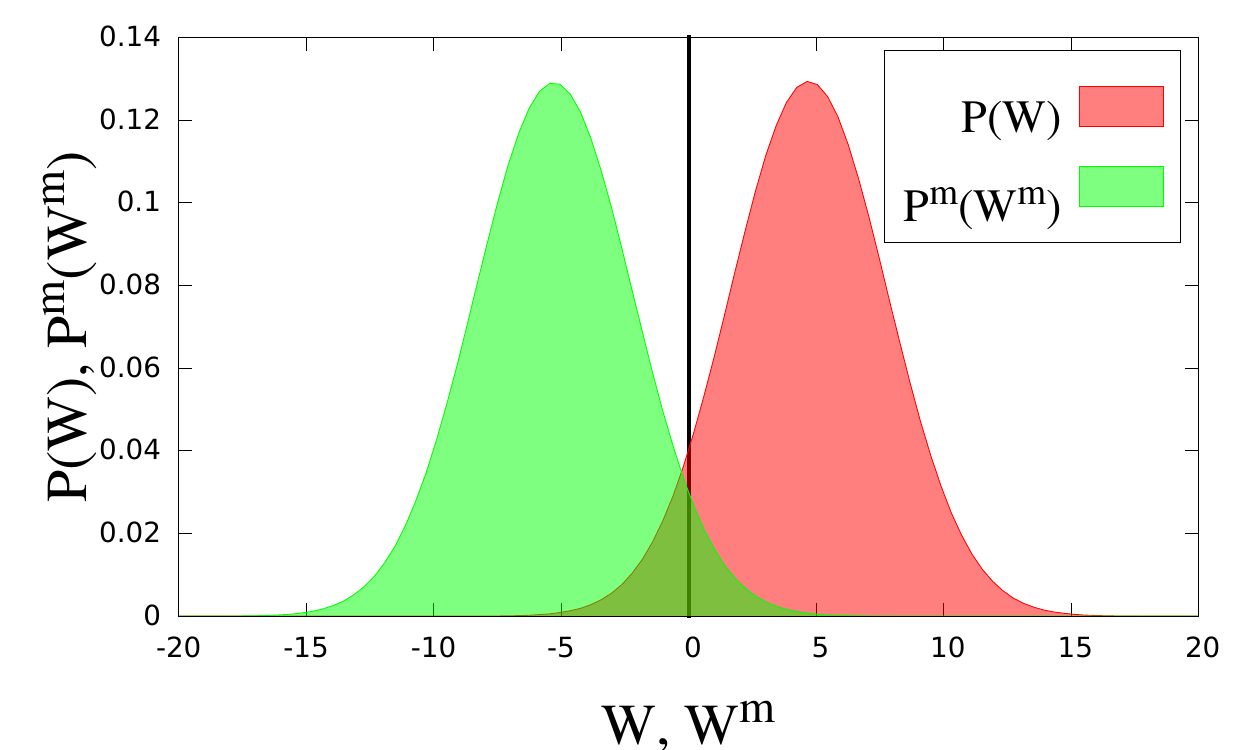}
  \caption{Plots of measured and true work distributions (work being measured in unit of $k_BT$), 
  when $\tau=20$ and $\av{\eta}>b/\kappa$. 
The parameters used are $b=0.5$, $\kappa=1$ and $\av{\eta}=1$.}
  \label{fig:viol}
\end{figure}

It is instructive to illustrate this result with a simple example. 
Consider a Brownian particle following the Langevin Equation
\begin{align}
  \dot x=-\kappa(x-\lambda)+\xi(t),
  \label{simple}
\end{align}
where $\xi(t)$ is the Gaussian random white noise: $\av{\xi(t)}=0$, and $\av{\xi(t)\xi(s)}=2T\delta(t-s)$. 
The system is initially at equilibrium with a heat bath at temperature $T$, and is 
thereafter perturbed by a time-dependent linear protocol $\lambda(t)=at/\tau\equiv bt$, where $b=a/\tau$.
The average work done on the particle is given by
\begin{align}
  \av{W} &= \int_0^\tau \dot\lambda(\lambda-\av{x})dt \nn\\
  &= b\int_0^\tau (bt-\av{x})dt.
  \label{avgwork}
\end{align}
Using Eqs. \eqref{simple} and \eqref{avgwork}, we arrive at
\begin{align}
  \av{W}&= \frac{b^2}{\kappa}\bigg[\tau-\frac{1-e^{-\kappa \tau}}{\kappa}\bigg].
\end{align}
With the present form of trapping potential, it is simple to check that 
the partition function is independent of $\lambda$ and as a result $\Delta F=0$. 
Thus, to allow $\meas{\mathfrak{f}}$ to be greater than 1/2, one should have $\av{\meas W}<0$, or equivalently 
$\varepsilon>\av{W}$, which means:
\begin{align}
  \varepsilon > \frac{b^2}{\kappa}\bigg[\tau-\frac{1-e^{-\kappa \tau}}{\kappa}\bigg].
  \label{condition_epsilon}
\end{align}
Using the inequality $e^r\ge 1+r$, it can be easily checked that the right hand side is always  non-negative.
If the measured position $\meas x$ and the true position $x$ are related by Eq.~(\ref{caseA})
assuming $\eta$ is another Gaussian distributed white noise, 
then the mean of $\eta$ is related to $\varepsilon$ as
\begin{align}
  \av{\eta} &= \frac{\varepsilon}{b\tau}.
\end{align}
Then the condition \eqref{condition_epsilon} translates to
\begin{align}
  \av{\eta} >\frac{b}{\kappa}\bigg[1-\frac{1-e^{-\kappa \tau}}{\kappa\tau}\bigg].
\end{align}
For large enough value of $\tau$, we then have the condition $\av{\eta}>b/\kappa$. 
When this condition is satisfied, we expect the mean of the Gaussian distribution of 
$\meas W$ to lie to the left of the $\meas W=0$ axis. This is shown in figure \ref{fig:viol}, 
where the distributions for the true work and the measured work have been plotted, 
with $b=0.5$, $\kappa=1$ and $\av{\eta}=1$. Clearly, the mean of the distribution 
for $\meas W$ lies to the left of the 
$\meas W=0$ axis, unlike for the distribution of the true work. 

This example shows that a sufficiently large and positive mean error drastically alters our estimation of the fraction 
of trajectories that violate the second law. Naturally, nothing of that sort would occur if $\varepsilon<0$.
Below, we test the main results of the paper regarding modified fluctuation theorems obtained in previous sections numerically. 
We start with the case of uncorrelated Gaussian error and then we consider an example of correlated non-Gaussian error. 

\subsection{Numerics for the case of uncorrelated Gaussian errors}
\label{sec:uncorrelated_Gaussian}
We begin by verifying the relation \eqref{amazing}, for a system that is subjected to the time-dependent potential
\begin{align}
  V(x,t)=\frac{1}{2}\kappa(x-\lambda(t))^2-\frac{1}{2}x^2+\frac{1}{4}x^4,
\end{align}
where the first term on the RHS represents the force acting on the system due to the harmonic trap, the center of which is 
positioned at $\lambda(t)$, and $\kappa$ is the stiffness constant of the trap. The second and third terms represent 
a double-well potential that the particle sees in addition to the trap potential. 
The system follows the overdamped Langevin equation of motion:
\begin{align}
  \dot x= -\partial_xV(x,t)+\xi(t),
\end{align}
$\xi(t)$ being the Gaussian thermal white noise with zero mean: $\av{\xi(t)}=0$, $\av{\xi(t)\xi(t')}=2T\delta(t-t')$. 

 We have chosen the parameter $\lambda(t)$ to be $A\sin\omega t$, which is a simple sinusoidal 
 drive of amplitude $A$ and frequency $\omega$. 
The protocol is applied for a time $\tau=\pi/2\omega$, which is one-fourth of the drive period. 
The error in the measurement corresponds to case $(a)$, with the additional assumption that $\eta$ is a Gaussian white 
noise of mean zero and of autocorrelation function $\av{\eta(t)\eta(t')}=\sigma_\eta^2 \delta(t-t')$. 
The error in the measurement of the work is also Gaussian, since it is linear in $\eta$:
\begin{align}
  E=\int_0^\tau dt \dot\lambda(t)\eta(t).
\end{align}

We can then derive the variance of the error to be
\begin{align}
  \sigma^2 &= \sigma_\eta^2\int_0^\tau dt\int_0^\tau dt'  \delta(t-t')\dot\lambda(t)\dot\lambda(t')\nn\\
  &= \sigma_\eta^2 A^2\omega^2 \int_0^\tau dt\cos^2\omega t \nn\\
  &= \sigma_\eta^2 A^2\omega\pi/4
\end{align}
for $\tau=\pi/2\omega$.
With the choice of parameters $\sigma_\eta^2=0.5$, $A=2$, and $\omega=1$, we obtain $\sigma^2\simeq 1.571$. 
The required shift in $\meas W$ is $R=\beta\sigma^2/2\simeq 0.785$.
 
\begin{figure}[!h]
    \centering
    \includegraphics[width=8cm]{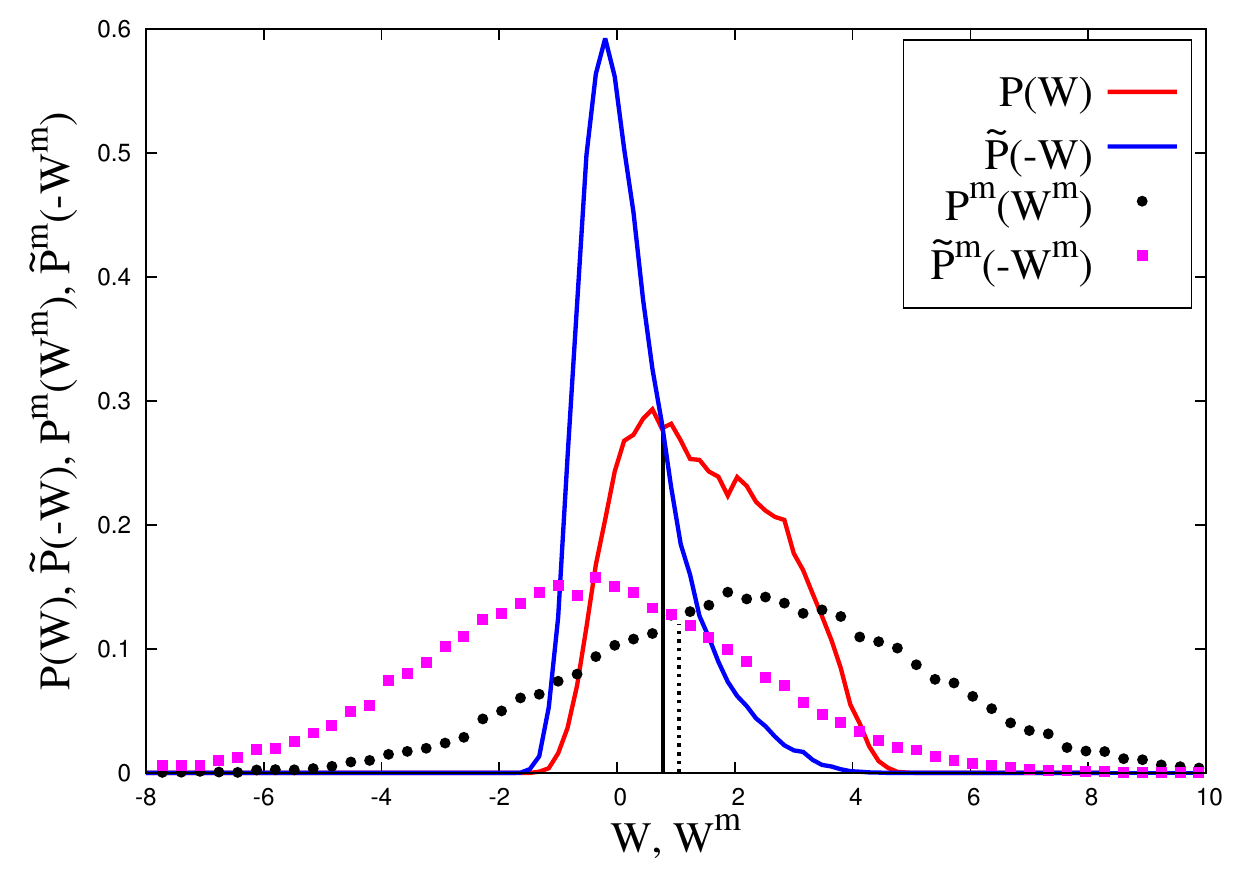}
    \caption{The distributions $P(W)$, $\tilde P(-W)$, $\meas{P}(W)$ and $\tmeas{P}(-W)$ for the above potential. 
    They are clearly non-Gaussian. The intersection point of $P(W)$ and $\tilde P(-W)$ gives the value of 
    $\Delta F\simeq 0.82$. The solid vertical line shows the value of $W$ at the intersection of $P(W)$ and $\tilde P(-W)$, whereas the dotted vertical line shows the value of $\meas W$ at the intersection of $\meas{P}(W)$ and $\tmeas{P}(-W)$. We have chosen $\sigma_\eta^2=2$, $A=2$, $k_BT=1$, $\kappa=1$, $\gamma=1$ and $\sigma_\eta^2=2$.}
    \label{fig:Wdist}
  \end{figure}

Figure \ref{fig:Wdist} shows the distributions of the true and the measured works for the above potential, 
for forward and reverse drivings. The parameters chosen have been mentioned in the figure caption. 
 The non-Gaussian nature of these distributions is apparent, as is the bias in the determination 
of free energy from Crooks relation. Indeed, the crossing point of $P(W)$ and $\tilde P(-W)$ (which gives the 
free energy change $\Delta F\simeq 0.82$) is clearly different from that of the distributions $P(\meas W)$ and 
$\tilde P(-\meas W)$. We also note that the variance of measured work in either the forward or the reverse 
process is higher than that of the true work. 

\begin{figure}[!h]
  \centering
  \includegraphics[width=8cm]{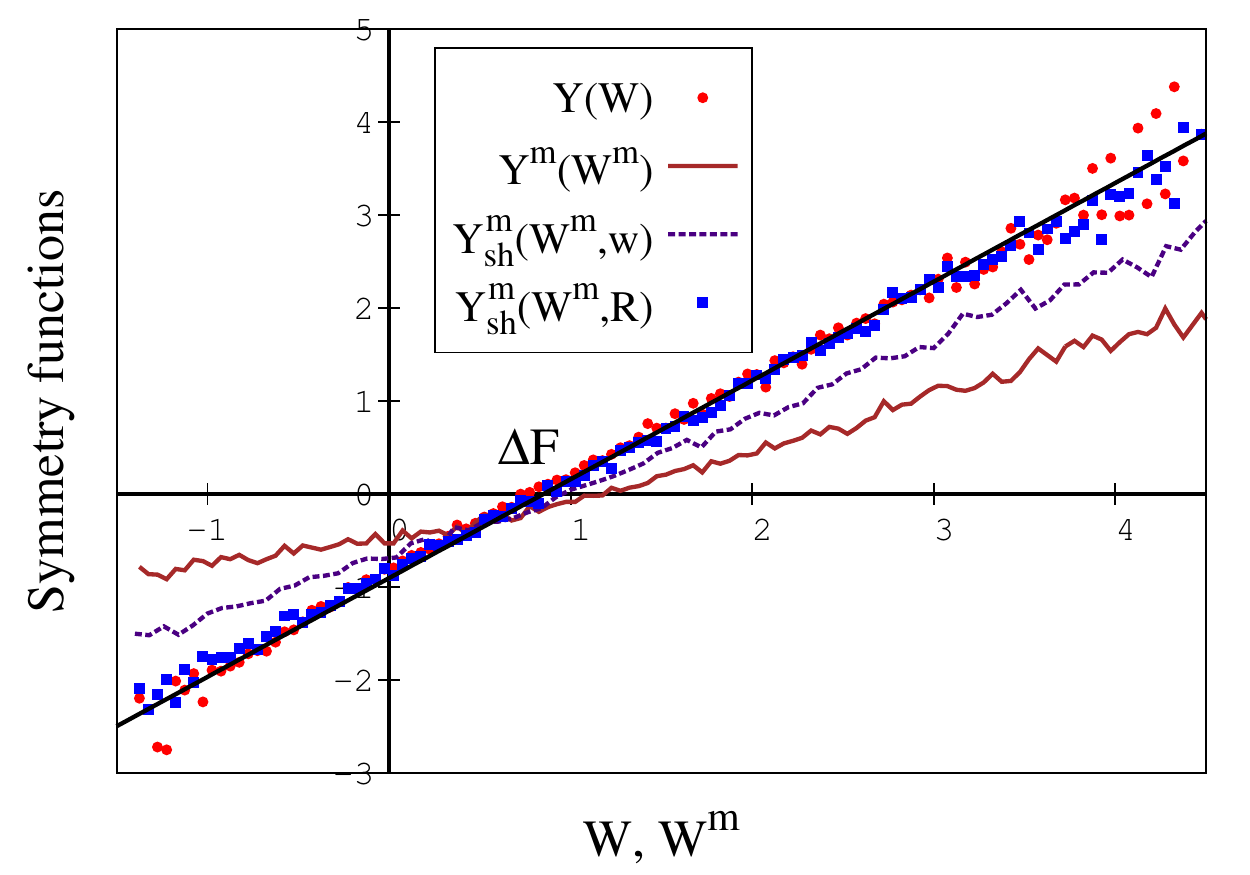}
  \caption{Symmetry functions for the true work $Y(W)$ (red filled circles), for 
  measured work $\meas Y(\meas W)$ 
  (brown solid line) and shifted symmetry functions of measured works $\meas Y_{\text{sh}}(\meas W,w)$
  with intermediate (dashed line) and optimal (blue filled squares) values of the shift $w$, as defined in 
  Eqs. \eqref{Y}, \eqref{Ym} and \eqref{Y_sh} respectively. 
  All energies are measured in units of $\beta^{-1}$. The black solid line is a linear fit for $Y(W)$. We have chosen $\sigma_\eta^2=0.5$, all other parameters being the same as in figure \ref{fig:Wdist}.
  The free energy, obtained from the Jarzynski equality is $\Delta F\simeq 0.82$.
  }
  \label{fig:sym}
\end{figure}

In figure \ref{fig:sym}, we show the symmetry functions for the true and the measured works, 
as a function of $W$, which are denoted by 
$Y(W)$ and $\meas Y(\meas W)$, respectively. The black solid line is the linear fit for $Y(W)$, 
which as expected corresponds to a straight line of slope one.  
In contrast, the symmetry function $\meas Y(\meas W)$ for the measured work is not a linear 
function as expected theoretically.

In view of the exact relations derived for shifted symmetry functions in Eqs. \eqref{correction-5}-\eqref{amazing}, we 
consider again shifted symmetry functions defined as in Eq.~\eqref{Y_sh}.
Initially the appropriate value of $w$ is unknown, but it is possible to tune this parameter so as to find the point 
where $w=R$. Here, since $\varepsilon=0$, the correct value is $R=\beta\sigma^2/2 \simeq 0.785$, at which point according to Eq. \eqref{amazing}, 
the shifted symmetry function has a simpler form, namely:  
\begin{align}
  \meas Y_{\text{sh}}(\meas W,R)=\beta(\meas W-\Delta F).
  \label{Y_sh2}
\end{align}

In fig.~\ref{fig:sym}, these shifted symmetry functions are shown for intermediate values of $w$ and for the 
most appropriate value $w=R$, which makes the data points collapse on the black line of slope 1. 
This shows that at least in this case where $\varepsilon=0$, 
it is indeed possible to tune the value of $w$ to infer the correct value of the free energy difference 
using only the noisy data of measured works. The correct value of the free energy in this simulation is $\Delta F \approx 0.82$, 
which corresponds well to the point where  the symmetry function of the true work and the shifted symmetry function  
of the measured work intersect the $W$-axis. In case we had $\varepsilon\neq 0$, we would be able to collapse the data on a straight line but would not be able to infer the correct $\Delta F$.

\subsection{Numerics for the case of correlated non-Gaussian error}
\label{sub-corr-err}
We now verify numerically our results obtained for non-Gaussian correlated error of the type 
$E=\uncore+\alpha\meas W$. 
This kind of error can arise in two situations: 
First, when there is a miscalibration of the trap stiffness used for 
the evaluation of the measured work, as discussed in subsection \ref{subsec:Jarzynksi}. 
Secondly, when the relation 
between the measured position and the true position is modified with respect to the one given by Eq.~\eqref{caseA}.  
It can be shown that the correct modification compatible with an error of the form $E=\uncore+\alpha\meas W$ is:
\begin{align}
\label{caseA2}
   \meas x = \frac{x+\alpha\lambda(t)+\eta}{1+\alpha},
\end{align}
where as before $\lambda$ denotes the position of the trap, 
while $\eta$ represents a random process which is uncorrelated
from $x(t)$.

\begin{figure}[!h]
    \centering
    \includegraphics[width=8cm]{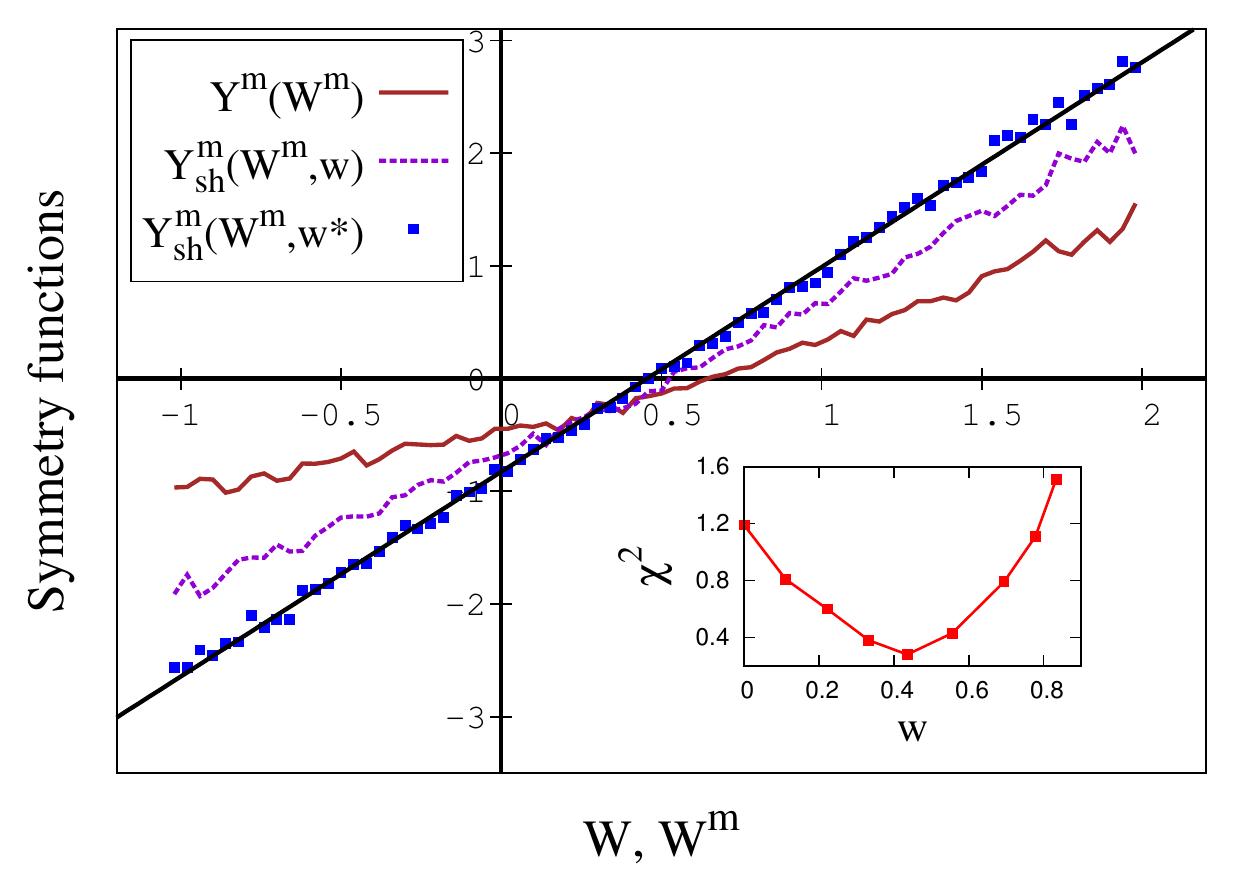}
    \caption{Symmetry functions for $\meas W$ without shift (brown solid line), intermediate shift $w=0.22$ (violet dashed line) and optimal shift $w^*=0.44$ (blue filled squares). 
    The black solid line is  the linear fit for the data points for optimal shift. The slope of this line is 1.82, which is very close to the actual value 1.8. The inset shows a plot of the $\chi^2$ of a linear fit of $\meas Y(\meas W,w)$ for various values of $w$. The minimum is reached at the optimal value $w=w^*$.
    Parameters: same as in figure \ref{fig:sym}.}
    \label{fig:sym_gen}
  \end{figure}

In our numerical simulations, we have implemented the second case for convenience.
In figure \ref{fig:sym_gen}, we have plotted the symmetry functions for the true and the measured 
works (filled circles and solid line in brown, respectively). Our parameters are the same as in 
figure \ref{fig:sym}. As in the case of figure \ref{fig:sym}, the symmetry function for $\meas W$ 
is not a linear function of $\meas W$. 

Improved free energy estimators can be constructed using shifted symmetry functions defined in Eq. \eqref{Y_sh} and by 
tuning the shift parameter $w$ until the data points collapse on a straight line. 
According to Eq. \eqref{amazing1}, the slope of the straight line at the collapse can be used to determine $\alpha$,
while the optimal value of $w$ provides information on $\uncorvar$ the variance of $\uncore$, since they are related by
 $w^*=\beta\uncorvar/(2(1+\alpha))$. 
From the linear fit of the data points for the optimal shift, we obtain a slope of $1.82$, 
which gives $\alpha=0.82$, in agreement with the actual value of 0.8.
In the inset of figure \ref{fig:sym_gen}, we show the plot for the $\chi^2$ values for this linear fit 
with the minimum of the plot coinciding with the optimal value $w^*\simeq 0.44$. This confirms that this
method can be used practically for determining $\alpha$ and $\uncorvar$.

\subsection{Correlated error due to finite delay in measurements}
\label{correlated_delay}

We next turn our attention to symmetry functions for measured work, when the measurement outcome is obtained after a time delay 
as described by case $(b)$ in Eq.~(\ref{caseB}). We rewrite the equations for $x$ and $\meas x$ below:
\begin{align}
  \gamma\dot x&= \kappa(\lambda(t)-x)+\xi(t); \nn\\
  \tau_r\meas{\dot x}&= x-\meas x+\eta(t).
  \label{delay}
\end{align}
Both, $\xi(t)$ and $\eta(t)$ are Gaussian white noises, of mean zero and of autocorrelation function $\av{\xi (t) \xi (t')}=2 \gamma T \delta(t-t')$ 
and $\av{\eta(t)\eta(t')}=\sigma_\eta^2\delta(t-t')$. 
We let the system evolve under these two equations in the absence of driving, $\lambda=0$, for a time much larger than $\gamma/\kappa$ as 
well as $\tau_r$, so that the true position variable $x$ is at equilibrium at time $t=0$, while the measured position $\meas{x}$ is 
in a non-equilibrium steady state. 

Due to the delay, the state variable $\meas x$ obeys a non-markovian dynamics, and as a result,
the error is correlated with the true work in a complex way. 
Linear Langevin systems with time delays have been studied extensively in the literature on feedback systems \cite{Munakata2014,Horowitz2014}. 
In that respect, our problem is simpler in that there is no feedback since the first equation in Eq. \eqref{delay} does not contain the variable 
$\meas{x}$.
Yet, the correlations between the error $E$ and the true work are more complex than that considered in Eq.~\eqref{alpha-case} in terms of the parameter $\alpha$. 
In particular, the error does not transform under time reversal as $\rho(E)=\tl\rho(-E)$ because 
the quantity $\err S$ which has been assumed to vanish in section \ref{sec:general} does not vanish here. 

{Fortunately, due to the linearity of the equations, the true and the measured works are also Gaussian distributed, 
being linear in $x$ and $\meas x$, respectively. 
Thus, we only need to focus on the mean and the variance of the measured work without 
having to consider the statistics of the error. 
The mean and the variance of the measured work can be obtained by direct integration of Eq.~\eqref{delay} 
after some algebra, which is detailed in the appendix 
\ref{app:meanvar}. 
From the formal expressions, one notices that the mean $\av{\meas W}$ as well as the variance 
$\sigma_{\meas W}^2$ are even under time-reversal symmetry, which implies $P(\meas W)=\tl P(\meas W)$. Using the fact that 
the distributions are Gaussian, one then arrives at the relation
\begin{align}
\label{slope}
  \meas Y(\meas W) &\equiv \ln\frac{P(\meas W)}{\tl P(-\meas W)}= \frac{2\meas W\av{\meas W}}{\sigma_{\meas W}^2}.
\end{align}
The slope of the symmetry function is therefore:
\begin{align}
\label{beta_eff} 
\beta_{eff} =\frac{2\av{\meas W}}{\sigma_{\meas W}^2}.
\end{align}
 It can be interpreted as an 
inverse effective temperature in view of the Crooks relation $\meas Y(\meas W) = \beta_{eff} \meas{W}$, which takes this form 
since the free energy difference is zero in the present setup.
\begin{figure}[!h]
  \centering
  \includegraphics[width=8cm]{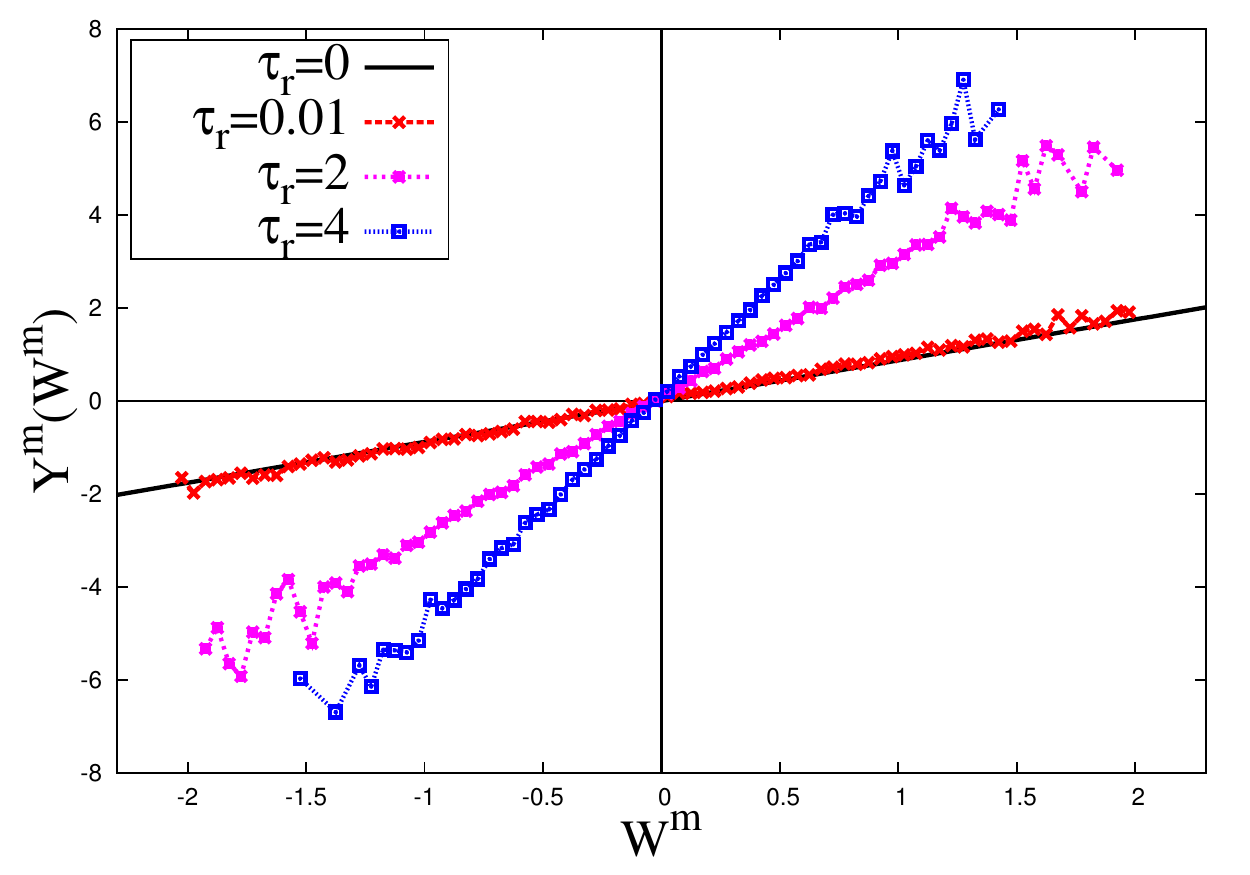}
  \caption{Symmetry functions $\meas Y(\meas W)$ for various values of the relaxation time $\tau_r$. 
The solid line is of slope $\sigma_W^2/(\uncorvar+\sigma_W^2)=0.878$, as expected when the error becomes uncorrelated with the work. 
The other parameters are: $a=1$, $k_B T=1$, $\sigma_\eta^2=0.1$ and $\tau=1$.}
  \label{fig:sym_delay}
\end{figure}
The symmetry functions for different values of $\tau_r$ for the applied linear protocol $\lambda(t)=at/\tau$ have been plotted in 
figure \ref{fig:sym_delay}. As expected all these curves are straight lines going through the origin, which confirms the 
interpretation in terms of effective temperatures. Note that this effective temperature depends on both 
$\tau_r$ and $\tau$ as we discuss now. 

As $\tau_r$ is increased, the inverse effective temperature increases, {\it i.e.} the effective temperature decreases. 
This is expected since by increasing $\tau_r$, the fluctuations of $\meas x$ become more and more smooth as a result of filtering
the fluctuations of the true position for longer measurement times. This filtering 
translates into a decrease of the fluctuations of the measured position, {\it i.e.} a decrease of their effective temperature. 
For $\tau_r\to 0$ (red curve), we recover the effective temperature, which has been obtained in Eq.~\eqref{fin} 
for the case of dynamics (a). That effective temperature is necessarily larger than the bath temperature and is shown by the black solid line. 

From the point of view of the measured position, the true position appears as a perturbation, or as a driving force which is imposed from the outside. 
This driving force imposes a new time scale $\gamma/\kappa$ on the dynamics of the measured position, 
which would evolve otherwise with the time scale $\tau_r$.  
According to Ref. \cite{Dieterich2015}, the regime for which an effective temperature is expected is the one for which 
  $\tau_r \ge \gamma/\kappa$, which corresponds to the case where we find a small effective temperature. Interestingly, we
  also have a well-defined effective temperature in the other regime  $\tau_r \le \gamma/\kappa$, which we have analyzed before.

\begin{figure}[!h]
  \centering
  \includegraphics[width=8cm]{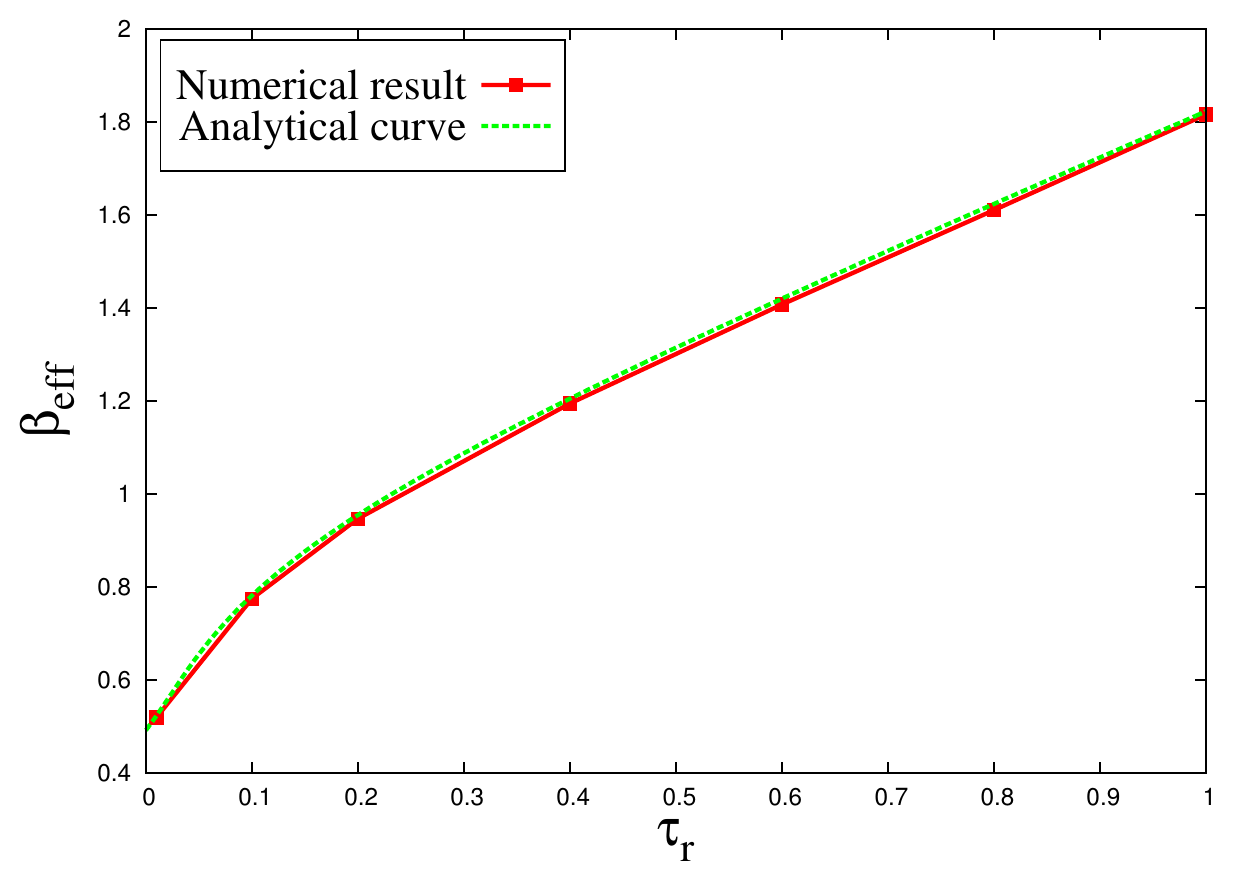}
  \caption{Inverse effective temperature $\beta_{eff}$ as a function of 
  the relaxation time $\tau_r$. 
  The parameters are $a=1$, $\tau=0.1$, $\sigma_\eta^2=0.1$, $k_B T=\kappa=\gamma=1$. 
  The green dotted line is an exact evaluation using Eq. \eqref{beta_eff} and appendix \ref{app:meanvar}, 
while the red solid line is the simulation curve.}
  \label{fig:tausmall}
\end{figure}

Naturally our case differs from that of Ref. \cite{Dieterich2015}, which is concerned with steady states, while we do not. 
This can also be seen from the dependence of our results on the time scale associated with the duration of the driving $\tau$. 
In figure \ref{fig:tausmall}, we show the variation of the slope of $\meas Y(\meas W)$, namely $\beta_{eff}$, as a function of $\tau_r$, 
when $\tau=0.1$. The other parameters are as mentioned in the figure caption.
 We find a very good agreement with the simulations in the full range of variation of $\tau_r$. 
 The same plots for higher observation time, $\tau=1$, is shown in figure \ref{fig:taularge}. 
Once again, there is a good agreement with the numerics, which in this case is a straight line of slope $\approx 0.95$.

In the limit $\tau_r\to\infty$, one can show from the exact expression of the slope of $\meas Y$ provided in appendix \ref{app:meanvar} that 
it behaves as $2\kappa \tau_r/(2\gamma T+\sigma_\eta^2\kappa^2)$, which correctly predicts the slope of the straight line in figure \ref{fig:taularge}.

\begin{figure}[!h]
  \centering
  \includegraphics[width=8cm]{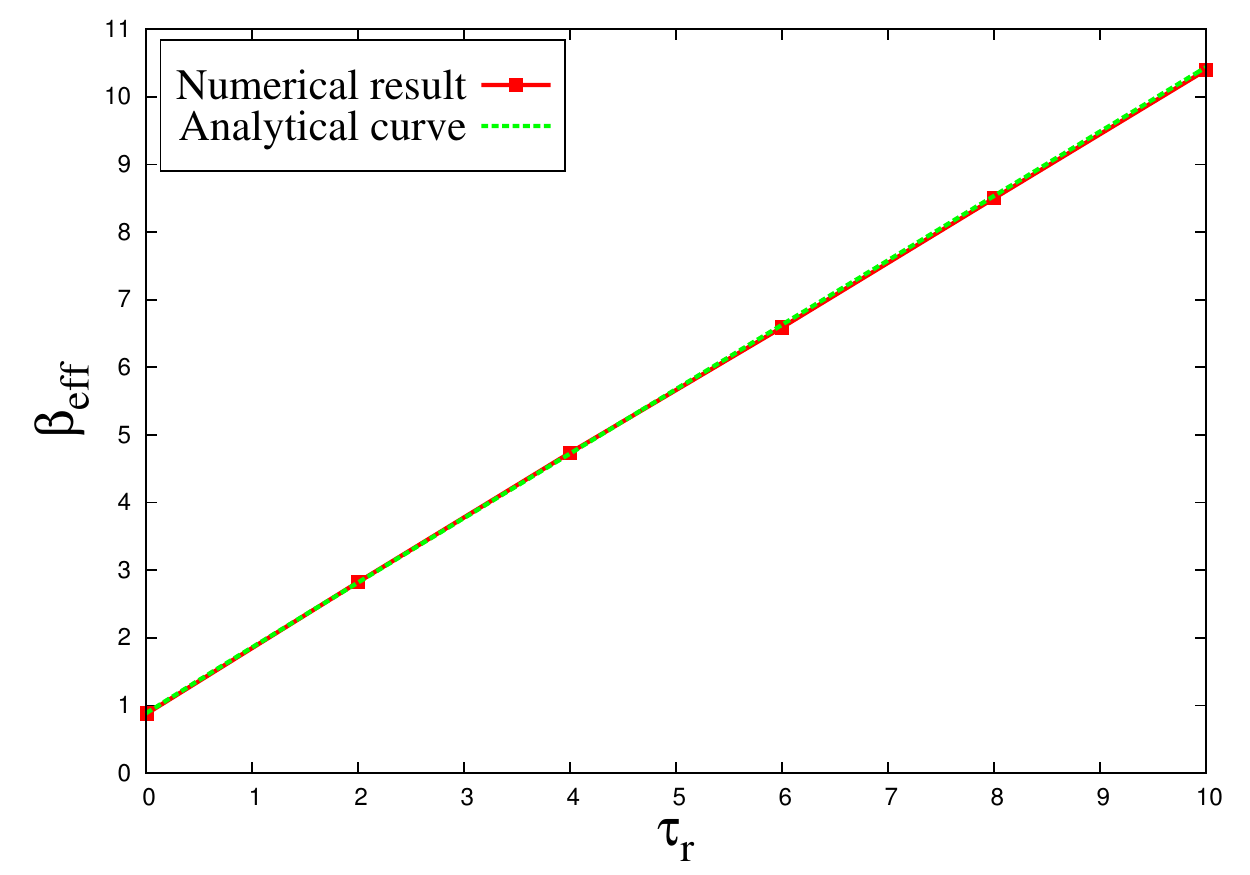}
  \caption{Same plot as above, with $\tau=1$, other parameters being the same.}
  \label{fig:taularge}
\end{figure}

\section{Relation to information theory and feedback}
\label{sec:feedback}
\subsection{Equilibrium initial conditions}

The corrections $\Omega$ and $R$ are factors modifying the fluctuation theorems in the presence of coarse-graining or noise.  
As shown below, similar factors have been introduced before in the context
of Fluctuation theorems with feedback \cite{Shiraishi2015,Sagawa2010_vol104}  and on related second-law like inequalities with information
\cite{Horowitz2014,Munakata2014}.

Let us consider the probability for a true trajectory in phase space, $\M{P}[X]$, and the one
for the measured trajectory $\meas{\M{P}}[\meas{X}]$. These two probabilities are related by
\begin{equation}
 \label{bias-work-1}
 \meas{\M{P}}[\meas{X}]=\int\M{D}X\err{\M{P}}[\meas{X}|X]\M{P}[X],
\end{equation}
where the difference between both distributions is contained in $\err{\M{P}}$.
In particular, when the measurement process is free of error,
$\err{\M{P}}[\meas{X}|X]=~\delta[\meas{X}-X]$, then 
$\meas{\M{P}}[\meas{X}]=\M{P}[\meas{X}]$.

In the experiment, only the coarse-grained trajectory $\meas{X}$ is available. 
The amount of information provided by $\meas{X}$ about the true trajectory $X$ is quantified by the mutual information
\begin{equation}
 \label{Mutual-I}
 I_x=\int\M{D}X\M{D}\meas{X}\M{P}[\meas{X},X]
 \ln\frac{\M{P}[\meas{X},X]}{\M{P}[X]\meas{\M{P}}[\meas{X}]},
\end{equation}
where $\M{P}[\meas{X},X]$ is the joint path probability of true and measured trajectories. In particular, if
$\meas{X}$ and $X$ are independent random variables, $I_x=0$,
implying that no information can be extracted on $X$ from the knowledge of
$\meas{X}$. Introducing the stochastic mutual information, $\M{I}_x[\meas{X},X]$, through
the relation
\begin{equation}
 \label{stochastic-I}
 \frac{\meas{\M{P}}[\meas{X}]}{\err{\M{P}}[\meas{X}|X]}=e^{-\M{I}_x[\meas{X},X]},
\end{equation}
we can write the mutual information simply as $I_x=~\langle\M{I}_x[\meas{X},X]\rangle\ge0$.
Then, combining Eqs. (\ref{stochastic-I})
and (\ref{path-FT}), we may write
\begin{equation}
 \label{almost-info}
 \meas{\M{P}}[\meas{X}]\tl{\M{P}}[\tl{X}]=\M{P}[\meas{X},X]
 e^{-\beta(W[X]-\Delta F)-\M{I}_x[\meas{X},X]}.
\end{equation}
It is worth noting that, in terms of the true work, a Jarzynski relation as for feedback
processes~\cite{Sagawa2010_vol104} follows immediately:
\begin{equation}
 \label{info-Jarzynski-true}
\big\langle e^{-\beta(W[X]-\Delta F)-\M{I}_x[\meas{X},X]}\big\rangle_\Lambda=1.
\end{equation}
Since the Jarzynski relation holds in terms of $W[X]$, the second law derived from Eq. (\ref{info-Jarzynski-true}),
$\beta(\langle W\rangle-\Delta F)\ge-I_x$ is uninformative, since $I_x\ge0$ while 
$\beta(\langle W\rangle-\Delta F)\ge0$ holds in this case. We shall thus turn instead towards 
a modified Jarzynski relation in terms of the measured work, which is experimentally accessible.

A key point is to recognize that the lack of knowledge on the system is represented by two
contributions. One is, of course, the error in the measurement of the true trajectory. The second is,
as stated above, the mutual information which quantifies, how much one can infer about the true trajectories from 
the measured ones. We can thus introduce a unified
quantity measuring both effects as
\begin{equation}
 \label{total-uncertainty}
 \Delta_x[\meas{X},X]=\beta E[\meas{X},X]+\M{I}_x[\meas{X},X].
\end{equation}
With this, and using Eq. (\ref{def-E}), Eq. (\ref{almost-info}) can be rewritten as
\begin{equation}
 \label{almost-info-1}
 \meas{\M{P}}[\meas{X}]\tl{\M{P}}[\tl{X}]=\M{P}[\meas{X},X]
 e^{-\beta(\meas{W}[\meas{X}]-\Delta F)-\Delta_x[\meas{X},X]}.
\end{equation}
We thus have after direct integration
\begin{equation}
 \label{info-Jarzynski}
\big\langle e^{-\beta(\meas W[\meas{X}]-\Delta F)-\Delta_x[\meas{X},X]}\big\rangle_\Lambda=1.
\end{equation}
 
Comparing the Jarzynski relation for feedback processes~\cite{Sagawa2010_vol104} with
Eq. (\ref{info-Jarzynski}), we see that there is a similar structure, despite the fact that 
there is no feedback in our case. An important
difference between both cases is that $\langle\Delta_x\rangle$ does not
have a definite sign, because there is no particular sign for the measurement error. 
However, if the distribution of the measurement errors has non-negative
mean, then $\langle\Delta_x\rangle$ is non-negative, since $I_x \ge 0$. 
It is also interesting to note, by simple inspection of Eq. (\ref{Jarzynski-exp}), 
that $\exp(\beta R)$ is the analog of the efficacy parameter
introduced in \cite{Sagawa2010_vol104} for feedback processes and denoted $\gamma$ in that reference. 

\subsection{Generalization to the case of nonequilibrium initial distribution}

In this sub-section only, we extend the results of previous sections to situations 
where the initial distribution of the true variable $x(0)$ is not an equilibrium one, but rather an 
arbitrary distribution. To emphasize this difference, let us now denote the corresponding 
full trajectory with a prime as $X'=\{x(t)\}_{t=0}^\tau$, to distinguish it from the trajectory which we had 
denoted $X$ so far. When the initial condition $x(0)$ is not an equilibrium one, 
the modified Crooks relation becomes \cite{Esposito2011_EPL,Lahiri2015_INJP}
\begin{align}
  \frac{\tilde P[\tilde X']}{P[X']} &= e^{-\beta(W-\Delta F)+\Delta D},
  \label{initial_noneq}
\end{align}
where $\Delta D\equiv D(x_\tau,\tau)-D(x_0,0)$, with
\begin{align}
  D(x_t,t) = \ln\frac{p(x_t,t)}{p^{eq}(x_t,t)}.
\end{align}

Note that this relation can equivalently be written as 
\begin{align}
  \frac{\tilde P[\tilde X']}{P[X']} &= e^{-\beta(W-\Delta F_{neq})}, 
  \label{initial_noneq}
\end{align}
if one introduces the non-equilibrium  free energy \cite{Esposito2011_EPL,Lahiri2009_PRE}
\begin{align}
  F_{neq}(x_t,t) &=E(x_t,t)-Ts(x_t,t)= F(t)+TD(x_t,t),
\end{align}
where $s(x_t,t)$ is the stochastic entropy defined by $s(x_t,t)\equiv -\ln p(x_t,t)$ \cite{Seifert-Review}.

Although the initial distribution of the forward or the backward process 
are assumed to be general non-equilibrium distributions, 
we still assume that the initial distribution 
of the backward process is the same as the final distribution 
reached in the forward process. Proceeding the same way as before, 
we arrive at the variant of the 
fluctuation relation
\begin{align}
  \av{e^{-\beta(\meas W-\Delta F)+\Delta D-\Delta_x}}_\Lambda=1,
\end{align}
which shows that deviations from the Jarzynksi relation appear due to both 
uncertainties present in the initial and final distribution (described by $D$) and 
in the trajectories themselves (described by $\Delta_x$).
Naturally, in the limit when the initial \emph{and} final distributions are at thermal equilibrium, 
$\Delta D=0$ and we get back Eq. \eqref{info-Jarzynski}.

\subsection{Error due to incorrect assumption of initial distribution}

Let us consider a special situation in which the error is only present in the initial distribution of the position. 
In other words, while at the time $t=0$ there is a difference between the true position and the measured position, 
afterwards, there is no error, we assume $\meas x(t)=x(t)$ for $t >0$. In this case, the measured position is discontinuous at $t=0$, 
as if it was undergoing a sudden quench due to $x(0)$. 
The true and measured works are related through the relation
\begin{align}
  \meas W = W+\lambda(0)(x(0)-\meas x(0)),
\end{align}
since the quench causes a change in internal energy of $\lambda(0)(x(0)-\meas x(0))$. 
From this equation, the error in work is given by $E=\lambda(0)(\meas x(0)-x(0))$. If this error is uncorrelated with the true work $W$, 
then the relation \eqref{exp-convol} again holds for this case. Furthermore $W$ and $E$ are Gaussian, 
then the entire analysis of sec. \ref{subsec:uncorr-Gaussian} goes through and we can define an effective temperature 
analogous to Eq. \eqref{fin} from the slope of the symmetry function of $\meas W$, which will be entirely due to 
the uncertainty about the initial condition.

\section{Conclusion}
\label{sec:conclusions}
In this paper, we have studied thermodynamic inference
from coarse-grained data or noisy measurements based on fluctuation theorems.
We have focused on measurements of stochastic work as in the Jarzynski or Crooks relations, although
much of the ideas discussed here would also apply to other quantities than stochastic work, involving for instance 
entropy production \citep{Alemany_Inference}.
We have distinguished two forms of errors, one which originates from the evaluation of the work itself, and another one
which originates from the inaccuracy in the knowledge of degrees of freedom which are needed to evaluate the work. We have shown that 
the thermodynamic inference problem is greatly simplified when the error made on the work is Gaussian 
and uncorrelated. Interestingly, when the work is Gaussian distributed, this problem can be reformulated in terms of an effective temperature, 
which captures the effect of noise or coarse-graining. 

On the practical side, for Gaussian uncorrelated errors of zero mean, 
a shift in the log-ratio of the probability distributions is able to collapse the measurements 
points on a straight line, thus providing a simple solution to 
the thermodynamic inference problem of free energy.
Remarkably, this strategy still works, when the error is of the form of Eq. \eqref{alpha-case} in which case it 
contains an uncorrelated Gaussian part. 
However, when the correlations between the work and the 
error are due to measurement delays, this simple strategy fails and the situation appears more complex. 
For that case, we have introduced a solvable model based on linear Langevin equations 
which includes measurements delays. We have analyzed the model theoretically by deriving 
its effective temperature and we have checked our analytical results using simulations.

Finally, we note that the modified fluctuation theorems used to construct improved 
estimators of free energy differences, take 
a form which is very similar to that found in problems with feedback. This connection appears 
quite promising to address future thermodynamic inference problems.
We hope that our work will stimulate further theoretical and experimental work in that direction.

\acknowledgments
D L would like to dedicate this paper to the friendship and memory of Maxime Clusel. 
Maxime has made various significant contributions to 
Statistical Physics including in the last years of his life, groundbreaking achievements in 
the field of quantum stochastic thermodynamics. 
Maxime was always supportive of other people's work, his kindness and humility was exemplary and will remain a source of inspiration 
for all of us.

Finally, D L would like to also acknowledge stimulating discussions with G. Verley, J. Guioth, and L. Peliti.
S L thanks the Institute of Complex Systems (ISC-PIF), the Region Île-de-France, and the Labex CelTisPhyBio (N ANR-10-
LBX-0038) part of the IDEX PSL (\textnumero ANR-10-IDEX-0001-02 PSL) for financial support.
RGG thanks J. A. Morin for stimulating discussions at IMDEA Nanociencia, Madrid, during 
the gestation of the first ideas of this work. 
RGG also acknowledges the financial support of the LabeX LaSIPS (ANR-10-LABX-0040-LaSIPS) 
managed by the French National Research Agency under the "Investissements d'avenir" program 
(\textnumero ANR-11-IDEX-0003-02).

\appendix

\section{Proof of the relation $\rho(E)=\tl\rho(-E)$}
\label{app:proof1}
The condition $\rho(E)=\tl{\rho}(-E)$, which we call the invariance of the error distribution under time reversal symmetry, 
can be derived as follows. First, when
$\err{\M{P}}$ is invariant upon time-reversal symmetry, Eq. (\ref{error-ft}) holds. Using Eq. (\ref{factorized form})
in (\ref{error-ft}), one has
\begin{align}
\ln\frac{\rho(E)}{\tl{\rho}(-E)}+\ln \frac{P(W+E)}{\tl{P}(-W-E)}=\beta(W+E-\Delta F).
\end{align}
On the other hand, the Crooks
relation for the true work distribution leads to $\ln P(W+E)/\tl{P}(-W-E)=\beta(W+E-\Delta F)$, giving the expected result
$\ln\rho(E)/\tl{\rho}(-E)=0$. It is important to notice that for time-reversal invariance of $\rho(E)$,
time reversal invariance of $\err{\M{P}}$ is necessary, but not sufficient. Statistical independence between
$W$ and $E$ is also needed.

Alternatively, the derivation can also be done at the level of the trajectories, within dynamics of type (a) as defined in Sec.~\ref{sec:tweezers}. 
Let us consider single molecule experiments done with harmonic traps, for which the error defined at the level 
of the work is only a functional of $X-\meas X$ as in Eq. (\ref{work-decomposition-harmonic}). 
Then, the statistical independence of $W$ from $E$ translates
into the independence of $X-\meas X$ from $X$. This implies that $\err{\M{P}}[ \meas X | X]$
is only a functional of $\pmb{\eta}=X-\meas X$, where $\pmb{\eta}$ denotes the trajectory  
$\{x(t)- \meas x(t) \}_0^{\tau}$. Thus,
\begin{align}
 \label{transform-path}
 \M{D}\meas X\err{\M{P}}[\meas X|X] &=\M{D} \pmb{\eta} \bigg|\frac{\partial \meas X}{\partial 
 \pmb{\eta}}\bigg|\M P_\eta[ \pmb{\eta} |X]\nn\\
 &=\M D \pmb{\eta} \M P_\eta[ \pmb{\eta}], 
\end{align}
where $|\partial \meas X/\partial {\pmb \eta}|$ denotes the Jacobian of the 
transformation, which is equal to one. It follows from this that
\begin{align}
  \rho(E) &=\int\M D X\M P[X]\int\M D \meas X \err{\M{P}}[\meas X|X] \delta(E-E[X,\meas X])\nn\\
  &= \int \mathcal DX\M P[X]\int\mathcal D \pmb{\eta} \M P_\eta[\pmb{\eta}] \delta(E-E[\pmb{\eta}])\nn \\
  &= \int \mathcal D \pmb{\eta} \tl{\M{P}}_\eta [\tl{\pmb{\eta}}] \delta(E+\tl E[\tl{\pmb{\eta}}])\equiv\tl{\rho}(-E).
\end{align}
In the second step,  we have used the property $E[X,\meas X]=E[\meas X-X]$ and
we changed variables to $\pmb{\eta}$ using \eqref{transform-path}. 
In the third step, we used the normalization property 
$\int\mathcal DX\M P[X]=1$ and the 
property that $\err P$ is invariant under time reversal, {\it i.e.} $\M P_\eta[\pmb{\eta}]=\tl{\M{P}}_\eta[\tl{\pmb{\eta}}]$. 
The change in sign in the error upon time reversal has also been used.
Notice that this derivation relies on dynamics (a) defined in Sec.~\ref{sec:tweezers}, 
and justifies \emph{a posteriori}
the approach used in Sec.~\ref{sec:general}. 

In contrast to this derivation, the property $\rho(E)=\tl\rho(-E)$ is not expected to
hold in the case of dynamics (b).

\section{Proof of Eq.~\eqref{uncore-symmetry}}
\label{app:proof2}

In the case of correlated non-Gaussian error, we have
\begin{equation}
 \label{err}
 E=\alpha\meas W+\uncore,
\end{equation}
and we still assume $\err{\M{P}}[\meas X|X]=\tl{\err{\M{P}}}[\tmeas X|\tl X]$ (which again is compatible
with dynamics (a)), such that $\Sigma(\meas W,E)=0$. Thus, we still can write:
\be
\label{joint-ft}
\ln\frac{\meas P(\meas W,E)}{\tmeas P(-\meas W,-E)}=\beta(\meas W+E-\Delta F).
\ee
We have shown in the main text, that the joint probability of the measured work and the error reads in this case
\begin{equation}
 \label{joint-form}
 \meas P(\meas W,E)=(1+\alpha)P(\meas W+E)\rho_u(E-\alpha\meas W).
\end{equation}
Correspondingly, we have
\begin{equation}
 \label{joint-form-reversed}
 \tmeas P(-\meas W,-E)=(1+\alpha)\tl P(-\meas W-E)\tl\rho_u(-E+\alpha\meas W),
\end{equation}
which implies
\begin{align}
 \label{proof-1}
 &\ln\frac{\meas P(\meas W,E)}{\tmeas P(-\meas W,-E)} \nn\\
 &=\ln\frac{P(\meas W+E)}{\tl P(-\meas W-E)}+
 \ln\frac{\rho_u(E-\alpha\meas W)}{\tl\rho_u(-E+\alpha\meas W)}\nn\\
 &=\beta(\meas W+E-\Delta F)+\ln\frac{\rho_u(E-\alpha\meas W)}{\tl\rho_u(-E+\alpha\meas W)},
\end{align}
where in the second step we have used the fluctuation theorem for the true work distribution.
Direct comparison between Eqs. \eqref{joint-ft} and \eqref{proof-1}, leads to
\begin{equation}
 \label{proof-2}
 \ln\frac{\rho_u(E-\alpha\meas W)}{\tl\rho_u(-E+\alpha\meas W)}=0,
\end{equation}
which, due to the arbitrariness of $E$, and $\meas W$, means
\begin{equation}
 \label{proof-3}
 \ln\frac{\rho_u(\uncore)}{\tl\rho_u(-\uncore)}=0,
\end{equation}
for any value of $\uncore$.

The connection with dynamics (a) can be more clearly seen at the level of trajectories.
Let us assume that the error $E$ is associated with a relation between the measured position and the true one
of the form given in Eq.~\eqref{caseA2}.
In this case, since $E=W-\meas W\equiv
k\int_0^\tau dt\dot{\lambda}(\meas x-x)$, we have:
\bea
 \label{err-alpha}
 \uncore[\meas X,X] &=& k \int_0^\tau dt \dot{\lambda}\big[\big(1+\alpha)\meas x-x-\alpha\lambda], \\ \nn
                    &=& k \int_0^\tau dt \dot{\lambda} \eta,
\eea
which makes the uncorrelated error a functional of $\pmb{\eta}$. In addition, we note the property 
$\uncore[\meas X,X]=-\tl \uncore[\tmeas X,\tl X]$.

Let us consider the distribution of the uncorrelated error, which is
\begin{align}
 \label{proof-trajectories}
 \rho_u(\uncore) &=\int\M DX\M P[X]\int\M D\meas X\err{\M{P}}[\meas X|X]\delta(\uncore-\uncore[\meas X,X])\nn\\
 &= 
 \int\M DX\M P[X]\int\M D\pmb \eta \M P_\eta[\pmb \eta]\delta(\uncore-\uncore[\pmb \eta])\nn\\
 &= \int\M D\tl{\pmb{\eta}} \tl{\M{P}}_\eta[\tl{\pmb{\eta}}]\delta(\uncore+\uncore[\tl{\pmb{\eta}}])\nn\\
 &\equiv\tl \rho_u(-\uncore),
\end{align}
where in the second step, we have used the relation Eq.~\ref{transform-path}. 
The latter equation, namely Eq.~\ref{transform-path} still holds in the present case 
since the Jacobian $|\partial \meas X/\partial {\pmb \eta}|$ now equals $(1+\alpha)^{-1}$,
while $\err{\M{P}}[\meas X|X] = (1+\alpha) \M P_\eta \big[\eta \big]$ 
so that $\alpha$ dependent factors cancels.
In the third step, the normalization condition $\int\M DX\M P[X]=1$ has been used together with
the change of variable $\pmb \eta\to\tl{\pmb{\eta}}$ and the symmetry properties of
$\uncore[\pmb \eta]$ and $P_\eta[\pmb \eta]$.

\section{Uniqueness of the linear form of the symmetry function}
\label{app:proof}

In this appendix, we prove that (i) there is a unique value of $w$ such that Eq. \eqref{amazing1} holds, 
and (ii) for no other value of $w$, the symmetry function $\meas Y_{\T{sh}}(\meas W,w)$ 
is a linear function of $\meas{W}$.

For the first point, we start from Eqs. \eqref{s-f} and \eqref{s-r} of the main text. 
Doing the change of variable $y=(1+\alpha)(\meas W-w)+\uncore$
under the integral sign in Eq. \eqref{s-f}, and $y=(1+\alpha)(\meas W+w)+\uncore$ in \eqref{s-r}, we can write:
\begin{widetext}
\begin{align}
 \label{u-1}
 \meas Y_{\T{sh}}(\meas W,w) &=-\beta\Delta F+\frac{2}{\uncorvar}(1+\alpha)^2w\meas W+
 \frac{2\varepsilon}{\uncorvar}(1+\alpha)w+
 \ln\frac{\int dy P(y)e^{-\frac{1}{2\uncorvar}(y-\varepsilon)^2+\frac{1}{\uncorvar}(1+\alpha)(\meas W-w)y}}
 {\int dy P(y)e^{-\frac{1}{2\uncorvar}(y-\varepsilon)^2+\frac{1}{\uncorvar}(1+\alpha)
 \big(\meas W+w-\beta\uncorvar/(1+\alpha)\big)y}}\nn\\
 &=-\beta\Delta F+\frac{2}{\uncorvar}(1+\alpha)^2w\meas W
 \frac{2\varepsilon}{\uncorvar}(1+\alpha)w+\Phi(\meas W;w),
\end{align}
\end{widetext}
where we have introduced the function parametrized by $w$, 
\begin{equation}
\label{family}
\Phi(x;w)=
\ln\frac{\int dy P(y)e^{-\frac{1}{2\uncorvar}(y-\varepsilon)^2+\frac{1}{\uncorvar}(1+\alpha)(x-w)y}}
 {\int dy P(y)e^{-\frac{1}{2\uncorvar}(y-\varepsilon)^2+\frac{1}{\uncorvar}(1+\alpha)
 \big(x+w-\beta\uncorvar/(1+\alpha)\big)y}}.
\end{equation}

For proving point (i), we need to show that $w^*=\beta\uncorvar/2(1+\alpha)$ is the only real value of $w$ 
such that $\Phi(x,w)\equiv0$ for all $x\in\mathbb{R}$. Let us assume that there exists $w_1\neq w^*$,
such that $\Phi(x,w_1)\equiv0$. This would imply then that for $w=w_1$, the numerator and the denominator in
Eq. (\ref{family}) are equal for all $x\in\mathbb{R}$, or equivalently, that
\begin{align}
 \label{u-3}
 2\int dy P(y) &e^{-\frac{1}{2\uncorvar}(y-\varepsilon)^2+\frac{1}{\uncorvar}(1+\alpha)
 \big(x-\beta\uncorvar/2(1+\alpha)\big)y}\times\nn\\
 &\times\sinh\bigg[\bigg(-w_1+\frac{\beta\uncorvar}{2(1+\alpha)}\bigg)y\bigg]\equiv0,
\end{align}

where the last equality is obtained by substracting the numerator and the denominator of (\ref{family}). 
Given the arbitrariness of $x$, Eq. (\ref{u-3}) implies that the integrand has to be zero, which implies that 
the hyperbolic sine identically vanishes, or that $w_1=\beta\uncorvar/2(1+\alpha)=w^*$, 
which contradicts our initial assumption.

Now let us prove the second point (ii), namely that for no other value of $w$, 
the symmetry function $\meas Y_{\T{sh}}(\meas W,w)$ 
is a linear function of $\meas{W}$. 
To prove this, we first note that for any $w\in\mathbb{R}$ arbitrarily fixed,
$\Phi(x;w)$ is bounded. This is so because it is a continuous function of $x$, 
and furthermore $\lim_{x\to-\infty}\Phi(x,w)=\lim_{x\to+\infty}\Phi(x,w)=0$ for any $w\in\mathbb{R}$.
Summarizing, we have the following three properties:
\begin{enumerate}
 \item There is only one value of $w$, say $w^*=\beta\uncorvar/2(1+\alpha)$, such that $\Phi(x,w^*)\equiv0$ 
 for all $x\in\mathbb{R}$,
 \item $\lim_{x\to\pm\infty}\Phi(x,w)=0$ for any $w\in\mathbb{R}$, and
 \item For any $w\in\mathbb{R}$, $\Phi(x;w)$ is bounded in $\mathbb{R}$.
\end{enumerate}

In order to make the point, we need to prove that there is no real $w$
such that $\Phi(x;w)=Ax+B$, with $A, B\in\mathbb{R}$ not simultaneously zero. 
It is important to note that $A$ and $B$ must not be both zero, because we would then have $\Phi(x;w)\equiv0$, which
is only possible for $w=w^*$, by virtue of Property 1.
Let us assume that, indeed, there exists $w_2\in\mathbb{R}$, such that $\Phi(x;w_2)=Ax+B$ with $A$ and
$B$ not simultaneously zero. Now, given that Property 3 holds for any $w\in\mathbb{R}$, it holds in particular for
$w=w_2$, which implies that $A\equiv0$, otherwhise $\Phi(x,w_2)$ would not be bounded. We are thus left with
$\Phi(x,w_2)=B$ for all $x$, with $B\neq0$. This means, in particular, that we have 
$\lim_{x\to\pm\infty}\Phi(x,w_2)=B$. But Property 2 is valid for any value of $w$, in particular for $w=w_2$, thus
we have $B=0$, which contradicts our initial assumption. This proves that, apart from $w^*=\beta\uncorvar/2(1+\alpha)$,
no other value of the shift $w$ can make the symmetry function a linear function of $\meas{W}$.

\section{Derivation of the expressions for mean and the variance of $\meas W$ in sec. \ref{correlated_delay}}
\label{app:meanvar}

\subsection{Characterization of the initial conditions}

We first note that from the second line of Eq. \eqref{delay}, we have,
\begin{align}
  \meas x(t) &= \meas x(0)e^{-t/\tau_r}+\frac{1}{\tau_r}\int_0^t dt_1[x(t_1)+\eta(t_1)]e^{-(t-t_1)/\tau_r}.
  \label{xm}
\end{align}
Since $\av{x(0)}=0$, we obtain that the initial condition of the measured position satisfies $\av{\meas x(0)}=0$.

To get the variance of the measured position at the initial time, we take the Fourier transform of Eq. \eqref{delay} to get
\begin{align}
  x(\omega) &=\frac{\xi(\omega)}{\kappa-i\omega\gamma};\nn\\
  \meas x(\omega) &= \frac{x(\omega)+\eta(\omega)}{1-i\omega\tau_r},
\end{align}
where the $x(t)$ is related to its Fourier transform $x(\omega)$ through
\begin{align}
  x(t) &= \int_{-\infty}^\infty x(\omega)e^{-i\omega t}d\omega.
\end{align}
Similar definition holds for $\meas x(\omega)$. 
Using the Wiener-Khinchin theorem, we have $\av{|\xi(\omega)|^2}=2\gamma T$ 
and $\av{|\eta(\omega)|^2}=\sigma_\eta^2$. One can then write,
\begin{align}
  \av{(\meas x(0))^2} &= \int \frac{d\omega}{2\pi}\av{|\meas x(\omega)|^2}\nn\\
  &= \int \frac{d\omega}{2\pi}\bigg[\frac{2\gamma T}{[(\omega\gamma)^2+\kappa^2]
  [1+(\omega\tau_r)^2]}+\frac{\sigma_\eta^2}{1+(\omega\tau_r)^2}\bigg].
\end{align}
We note that the integrand has poles at $\omega=\pm i\kappa/\gamma$ and at $\pm i/\tau_r$. 
Choosing to integrate over the upper half 
of the complex plane, and using the fact that $\av{\meas x(0)}=0$, the variance of $\meas x(0)$  is
\begin{align}
  \sigma_{\meas x}^2(0) &=\av{(\meas x(0))^2} = \frac{\gamma T}{\kappa(\gamma+\kappa\tau_r)}+\frac{\sigma_\eta^2}{2\tau_r} .
\end{align}
By the same method, one also obtains the correlation function between the true and measured positions at the initial time, 
$\av{\meas x(0)x(0)}$ using again Fourier transforms. We find
\begin{align}
  \av{\meas x(0)x(0)}=\frac{\gamma T}{\kappa(\kappa \tau_r+\gamma)}.
\end{align}
With these expressions, we can proceed to calculate the mean and the variance of the measured work.

\vspace{1cm}

\subsection{Computation of $\av{\meas W}$}

From Eq. \eqref{xm}, we have
\begin{align}
  \av{\meas x(t)}=\frac{1}{\tau_r}\int_0^t dt_1\av{x(t_1)}e^{t_1/\tau_r}.
\end{align}
On the other hand, from \eqref{delay} we also have
\begin{align}
  x(t_1) &= x(0)e^{-\kappa t_1/\gamma}+\frac{1}{\gamma}\int_0^{t_1} dt_2[\kappa\lambda(t_2)+\xi(t_2)]
  e^{-\kappa (t_1-t_2)/\gamma}.
\end{align}
Combining the above two equations, it follows that
\begin{align}
  \av{\meas x(t)} &= \frac{\kappa}{\gamma \tau_r}\int_0^t dt_1 e^{-(t-t_1)/\tau_r}\int_0^{t_1}dt_2
  e^{-\kappa (t_1-t_2)/\gamma}\lambda(t_2).
\label{xmavg}
\end{align}
Now, 
\begin{align}
  \av{\meas W} &= \kappa\int_0^\tau dt\dot\lambda(t)(\lambda(t)-\av{\meas x(t)}).
\end{align}
We have chosen $\lambda(t)=at/\tau$. Thus, using \eqref{xmavg}, the following formal expression 
for $\av{\meas W}$ is obtained:
\begin{align}
  \av{\meas W} &= \frac{a^2\kappa}{2}-\frac{a^2}{\tau^2}\frac{\kappa^2}
  {\gamma\tau_r}\int_0^\tau dt \int_0^t dt_1 e^{-(t-t_1)/\tau_r}\nn\\
  &\times\int_0^{t_1}dt_2~t_2e^{-\kappa (t_1-t_2)/\gamma}.
\end{align}
This leads to the following expression for the mean measured work:
\begin{align}
  \frac{a^2 \left[\gamma ^3 \left(e^{-\frac{\kappa  \tau }{\gamma
   }}-1\right)+\gamma ^2 \kappa  \tau +\kappa ^3 \tau_r^2
   \left(-\tau_r e^{-\frac{\tau }{\tau_r}}-\tau
   +\tau_r\right)\right]}{\tau^2\kappa(\gamma -\kappa  \tau_r)}.
\end{align}

\subsection{Computation of $\sigma_{\meas W}^2$}

One can readily obtain the formal expression for $\sigma_{\meas W}^2$ in terms of measured position as:
\begin{align}
  \sigma_{\meas W}^2 &= \kappa^2\int_0^\tau dt\dot\lambda(t)\int_0^\tau dt'\dot\lambda(t')
  \av{\Delta\meas x(t)\Delta\meas x(t')},
  \label{Wvar}
\end{align}
where $\Delta \meas x(t)\equiv \meas x(t)-\av{\meas x(t)}$. Thus, we first need to calculate 
the quantity $\av{\Delta\meas x(t)\Delta\meas x(t')}$. We first note that
\begin{align}
  \Delta \meas x(t) &= \meas x(0)e^{-t/\tau_r}+x(0)\frac{t_c}{\tau_r}\big(e^{-t/\tau_r}-e^{-\kappa t/\gamma}\big)\nn\\
  &+\frac{1}{\gamma\tau_r}e^{-t/\tau_r}\int_0^t dt' e^{-t'/t_c}\int_0^{t'}dt_1 \xi(t_1)e^{\kappa t_1/\gamma}\nn\\
  &+\frac{1}{\tau_r}\int_0^t dt' \eta(t') e^{-(t-t')/\tau_r},
\end{align}
where $1/t_c=\kappa/\gamma-1/\tau_r$.
Thus, we have
\begin{widetext}
\begin{align}
  \av{\Delta\meas x(t)\Delta\meas x(t')} &= \av{(\meas x(0))^2}e^{-(t+t')/\tau_r} 
  + \av{x^2(0)}\bigg(\frac{t_c}{\tau_r}\bigg)^2\big(e^{-t/\tau_r}-e^{-\kappa t/\gamma}\big)
  \big(e^{-t'/\tau_r}-e^{-\kappa t'/\gamma}\big)\nn\\
  &+ \av{\meas x(0) x(0)}\frac{t_c}{\tau_r}\big[2e^{-(t+t')/\tau_r}-e^{-t/\tau_r-\kappa t'/\gamma}-
  e^{-t'/\tau_r-\kappa t/\gamma}\big]\nn\\
  &+ \frac{1}{\gamma^2\tau_r^2}e^{-(t+t')/\tau_r}\int_0^t dt_2e^{-t_2/t_c}\int_0^{t_2}dt_1
  e^{\kappa t_1/\gamma}\int_0^{t'}dt_3e^{-t_3/t_c}\int_0^{t_3}dt_4 e^{\kappa t_4/\gamma}\av{\xi(t_1)\xi(t_4)}\nn\\
  &+\frac{1}{\tau_r^2}\int_0^t dt_1 e^{-(t-t_1)/\tau_r}\int_0^{t'}dt_2 e^{-(t'-t_2)/\tau_r}\av{\eta(t_1)\eta(t_2)}.
\end{align}
\end{widetext}
The fifth term (fourth line) can be readily calculated to be
\begin{align}
  \frac{\sigma_\eta^2}{2\tau_r}\big[e^{-|t-t'|/\tau_r}-e^{-(t+t')/\tau_r}\big].
\end{align}
The fourth term (third line) can also be explicitly calculated. Finally, plugging these expressions into Eq. \eqref{Wvar}, 
we obtain the variance of the measured work. The explicit expressions are lengthy and not very illuminating, 
and for that reason are not given here.

\bibliographystyle{apsrev4-1.bst}
\bibliography{error}

\end{document}